\documentclass[aps,prd,secnumarabic,nobalancelastpage,amsmath,amssymb,
nofootinbib,twocolumn,showpacs]{revtex4-1}

\usepackage[toc,page]{appendix}
\usepackage{graphics}      
\usepackage{graphicx}      
\usepackage{longtable}     
\usepackage{url}           
\usepackage{bm}            
\usepackage{mathrsfs}
\usepackage{hyperref}
\usepackage{color}
\usepackage{soul}
\usepackage{tikz}
\usetikzlibrary{calc}

\numberwithin{equation}{section}
\newcommand{\secref}[1]{\S\ref{#1}} 
\newcommand{\appref}[1]{Appendix~\ref{#1}} 
\newcommand\ringring[1]{%
  {
   \mathop{\kern0pt #1}\limits^{
     \vbox to-1.85ex{
       \kern-2ex 
       \hbox to 0pt{\hss\normalfont\kern.1em \r{}\kern-.45em \r{}\hss}%
       \vss 
     }
   }
  }
}
\begin{document}
\title{Continuous body dynamics and the Mathisson-Papapetrou-Dixon equations}
\author{S. P. Loomis}
\email{sloomis@ucdavis.edu}
\altaffiliation[Current Address: ]{Department of Physics, University of California, Davis, California 95616}
\affiliation{Department of Physics, North Carolina State University, Raleigh, NC 27695}
\author{J. David Brown}
\email{david{\_}brown@ncsu.edu}
\affiliation{Department of Physics, North Carolina State University, Raleigh, NC 27695}
\date{\today}

\begin{abstract}
We show that an effective particle Lagrangian yields the Mathisson-Papapetrou-Dixon (MPD) equations. The spin of the effective particle is defined without any reference to a fixed body frame or angular velocity variable.
We then demonstrate that a continuous body, defined by a congruence of world lines and described by a general action, can be rewritten as an effective particle. We analyze the gauge freedom of the body and show that a natural center of mass condition is related to a spin supplementary condition.
\end{abstract}
\maketitle\tableofcontents
\makeatletter
\let\toc@pre\relax
\let\toc@post\relax
\makeatother


\section{Introduction}\label{introduction}
The motion of extended bodies in general relativity was first addressed by Mathisson in 1937 \cite{Mathisson-English}. Mathisson defined multipole moments for the stress-energy tensor expanded about a central world line and formulated the conservation of stress-energy as a variational principle. He derived what we now know as the Mathisson-Papapetrou-Dixon (MPD) equations to 
``pole--dipole"" order, and identified the quadrupole terms that had a nonrelativistic analogue. Papapetrou, using a different definition for multipole moments, derived the same equations in 1951 \cite{Papapetrou}. 

The analyses of Mathisson and Papapetrou yield ten equations for thirteen unknowns. The ten equations give the time evolution of the 
four components of momentum, $P^\alpha$, and the six components of 
spin, $S^{\alpha\beta}$. The thirteen unknowns are the momentum, spin, and the three degrees of freedom contained in the particle's proper velocity $U^\alpha$. 

Mathisson and later Pirani \cite{Pirani}  addressed the mismatch between the number of equations and number of unknowns 
by introducing the spin supplementary condition $U^\alpha S_{\alpha\beta}=0$. Pirani justified this choice by analogy to a similar identity for the center of mass in special relativity.
Papapetrou instead employed the spin supplementary condition $V^\alpha S_{\alpha\beta}=0$ where $V^\alpha$ is an arbitrary time flow vector field. In 1959 Tulczyjew simplified Mathisson's multipole formulation, again deriving the same equations of motion at pole-dipole order but choosing the spin supplementary condition $P^\alpha S_{\alpha\beta} = 0$,  arguing that Mathisson and Pirani's condition did not uniquely determine the world line. A survey of the various spin supplementary conditions and how they relate to one another can be found in \cite{Kyrian}. A  concrete analysis of the relationship between spin supplementary conditions and center of mass can be found in \cite{Costa15}. 

In a series of papers from 1970 to 1974, Dixon presented yet another reformulation of the multipole moments in terms of a Fourier transformation of the stress-energy tensor \cite{Dixon70-1,Dixon70-2,Dixon74}. The complete argument is also given in \cite{Dixon15}. Dixon found that Mathisson's variational principle yields dynamical equations for $P_\alpha$ and $S_{\alpha\beta}$, but leaves the dynamical evolution of the quadrupole and higher order multipole moments undefined. His analysis places restrictions on the symmetries of these multipoles. Based on these symmetries, 
Dixon defines the reduced multipole moments $J^{\mu\nu\rho\sigma\alpha_1\cdots\alpha_n}$ for $n\geq 0$.

The final form of the MPD equations through quadrupole order, as given by Dixon, is \cite{Dixon15}:
\begin{subequations}\label{MPD}
\begin{gather}\label{MPDp}
\frac{\mathrm{D}P_\alpha}{\mathrm{D}s}= -\frac{1}{2} R_{\alpha\beta\mu\nu}\dot{X}^\beta S^{\mu\nu} - \frac{1}{6}\nabla_\alpha R_{\mu\nu\rho\sigma}J^{\mu\nu\rho\sigma} \ , \\\label{MPDS}
\frac{\mathrm{D}S_{\alpha\beta}}{\mathrm{D}s} = 2P_{[\alpha}\dot{X}_{\beta]} + \frac{4}{3}R_{\mu\nu\lambda[\alpha}J_{\beta]}{}^{\lambda\nu\mu} \ .
\end{gather}
\end{subequations}
Here, the worldline is expressed as $x^\alpha = X^\alpha(s)$, where $x^\alpha$ are the spacetime coordinates and $X^\alpha$ are functions of a worldline parameter $s$.
The dot above a symbol denotes the time derivative $\mathrm{d}/\mathrm{d} s$ and $\mathrm{D}/\mathrm{D}s$ is the covariant derivative along the worldline; for example, $\mathrm{D}P_\alpha/\mathrm{D}s = \dot{P}_\alpha - \Gamma^\gamma_{\alpha\beta}\dot{X}^\beta P_\gamma$. This will also be denoted with a circle above the symbol, so that $\mathring{P}_\alpha = \mathrm{D}P_\alpha/\mathrm{D}s$.

Dixon defines the momentum and spin in terms of integrals over the leaves of a foliation $\Sigma(s)$ of the world tube. The integrals,  which involve the stress-energy tensor $T^{\mu'\nu'}$, Synge's world function $\sigma$ and the Jacobi propagators $H^{\mu'}{}_{\alpha}$ and $K^{\mu'}{}_{\alpha}$ (described in Appendix B) are \cite{Dixon70-1}
\begin{subequations}\label{DixonSnP}
\begin{align}\label{Dixonp}
P_{\alpha}(s) &\equiv \int_{\Sigma(s)}\mathrm{d}\Sigma_{\nu'}K^{\mu'}{}_{\alpha} T_{\mu'}^{\nu'} \ , \\\label{DixonS}
S_{\alpha\beta}(s) &\equiv 2 \int_{\Sigma(s)}\mathrm{d}\Sigma_{\nu'}H^{\mu'}{}_{[\alpha}\sigma_{\beta]} T_{\mu'}^{\nu'} \ .
\end{align}
\end{subequations}
Both $H^{\mu'}{}_{\alpha}$ and $K^{\mu'}{}_{\alpha}$ are bitensors with one primed index ``located'' at the point of integration $x^{\mu'}\in\Sigma(s)$ and one unprimed index located at the world line $X^\alpha(s)$. These bitensors act to transfer vectors from one tangent space into the other. The derivative of Synge's world function $\sigma_\beta = \nabla_\beta\sigma$ points opposite to the vector tangent to the unique geodesic connecting $X^\alpha(s)$ and $x^{\mu'}$ and acts as a position vector. The corresponding integrals for the multipole moments are complicated and  beyond the scope of this introduction.

The multipole method of Mathisson, Papapetrou, Tulzcyjew and Dixon is not the only approach that yields the MPD equations. Bailey and Israel \cite{Bailey}, extending the work 
of Hanson and Regge \cite{Hanson}, showed that a form of
the MPD equations  could be derived from any reparametrization--invariant Lagrangian involving a world line $X^\alpha(s)$, a set of Lorentz--orthonormal basis vectors $ e_a{}^\alpha(s)$ transported along the world line, and a set of external tensor fields that interact with the body. The basis vectors $ e_a{}^\alpha(s)$ define the orientation of the body, although the exact relationship for a 
physical ({\em i.e.} non--rigid) body is not addressed. By analogy with 
rigid body motion in classical mechanics, one says that the index $a$ labels the legs of a ``body--fixed frame''.

These results were independently replicated without the external fields by Porto \cite{Porto}{. More recently, Steinhoff \cite{Steinhoff}  has}  reformulated Bailey and Israel's full result in newer notation. We give a brief overview of Steinhoff's presentation below.

Steinhoff begins by considering an action of the form 
\begin{widetext}
\begin{equation}\label{SteinhoffAction}
S[X, e,\Phi_I] =
\int\mathrm{d}s\, L\left(g_{\alpha\beta}(X),\Phi_A(X),\dot{X}^\alpha(s), e_a{}^\alpha(s),\Omega^{\alpha\beta}(s),
\Phi_I(s)\right) \ ,
\end{equation}
\end{widetext}
where the integration is along the world line $X^\alpha(s)$.
Here $g_{\alpha\beta}(x)$ is the metric of the spacetime manifold $\mathscr{M}$, $\Phi_A(x)$ are fields on this manifold and $\Phi_I(s)$ are scalar dynamical variables
defined along the world line. Note that the $\Phi_I$ are functions of $s$ only, and the index $I$ can include a body frame index $a$.
The angular velocity is defined in terms of the basis vectors  $e_a{}^\alpha(s)$ by $\Omega^{\alpha\beta}(s) \equiv \eta^{ab} e_a{}^\alpha  \mathrm{D}( e_b{}^\beta)/\mathrm{D}s$, where $\eta^{ab}$ is the Minkowski metric. 

Steinhoff defines the momentum and spin as
\begin{subequations}\label{SteinhoffpS}
\begin{gather}\label{Steinhoffp}
P_\alpha \equiv \frac{\partial L}{\partial \dot{X}^\alpha} \ ,\\\label{SteinhoffS}
S_{\alpha\beta} \equiv 2\frac{\partial L}{\partial \Omega^{\alpha\beta}} \ .
\end{gather}
\end{subequations}
With these definitions, variation of the action \eqref{SteinhoffAction} with respect 
to $X^\alpha(s)$ and $ e_a{}^{ \alpha}(s)$ yields the equations of motion
\begin{subequations}\label{SteinhoffMPD}
\begin{gather}\label{SteinhoffMPDp}
\frac{\mathrm{D}P_\alpha}{\mathrm{D}s} = -\frac{1}{2}R_{\alpha\beta\mu\nu}\dot{X}^\beta S^{\mu\nu} + (\nabla_\alpha\Phi_A)\frac{\partial L}{\partial\Phi_A} \ , \\\label{SteinhoffMPDS}
\frac{\mathrm{D}S_{\alpha\beta}}{\mathrm{D}s} = 2P_{[\alpha}\dot{X}_{\beta]} - 2(\mathrm{G}_{[\alpha\beta]}\Phi_A)\frac{\partial L}{\partial\Phi_A} \ .
\end{gather}
\end{subequations}
Here, $\mathrm{G}_{\alpha\beta}$ is a linear operator defined such that $\nabla_\alpha \Phi_A = \partial_\alpha \Phi_A + \Gamma^\gamma_{\alpha\beta}\mathrm{G}^{\beta}{}_{\gamma}\Phi_A$. The action 
of $\mathrm{G}^\alpha{}_\beta$ depends on the tensor type of $\Phi_A$. 

In \secref{mpd} we consider a general action that { depends only on a worldline $x^\alpha = X^\alpha(s)$, a set of tensors $\psi_I(s)$ defined along the worldline, and a set of external fields $\phi_A(x)$.  We call this the ``effective particle'' model, and show that it 
yields the MPD equations.\footnote{After obtaining this result, we became aware of a similar result by H. Fuchs \cite{Fuchs}.} Our action} does not depend on an orthonormal 
body frame $ e_a{}^\alpha$ or angular velocity $\Omega^{\alpha\beta}$. We show in \secref{continuous} that continuous bodies (defined as a congruence of world lines minimizing a particular action) can be expressed in terms of the effective particle action. The definitions  for 
momentum and spin that emerge from this analysis coincide with the definitions given by Dixon \cite{Dixon70-1}. In \secref{conditions} we  discuss the gauge constraints that can be placed on the effective particle, and explore the relation between spin supplementary condition and center of mass. Finally, in \secref{example} we {apply our results to analyze a continuous body of noninteracting particles---a ``dust cloud." As an effective particle, the dust cloud satisfies the geodesic deviation equations.}

\section{The MPD Equations for an effective particle}\label{mpd}

In this section we generalize the method for deriving the MPD equations used by Bailey and Israel \cite{Bailey}. 

The system consists of an effective particle in a manifold $\mathscr{M}$ with position $x^\alpha = X^\alpha(s)$ and a collection of tensors $\psi_I(s)$. The 
tensors $\psi_I$ take the place of the orthonormal basis $ e_a{}^\alpha$ and 
body--frame variables $\phi_I$ used in Eq.~(\ref{SteinhoffAction}). 
The index $I$ can denote tensor indices as well as functional dependence. The evolution of the effective particle system is described by the action
\begin{equation}\label{EffectiveAction}
S[X,\psi_I] = \int_{s_0}^{s_1}\mathrm{ds}\,L(\phi_A(X),\dot{X}^\alpha(s),\psi_I(s),\mathring{\psi}_I(s)) \ .
\end{equation}
The { $\phi_A(x)$} are any collection of spacetime fields. 
For example, $\phi_A$ can include the electromagnetic field and its derivatives, the 
metric tensor $g_{\alpha\beta}$, the curvature tensor $R_{\alpha \beta\gamma\delta}$ and its symmetrized derivatives $\nabla_{(\alpha}\cdots\nabla_{\gamma)}R_{\mu \nu\rho\sigma}$. 
In this paper we treat these fields as external sources---they are not varied 
in the variational {principle. 

Recall} that the notation $\mathring{\psi}_I$ is an abbreviation for the covariant derivative along the worldline, 
\begin{equation}
\mathring{\psi}_I\equiv \frac{\mathrm{D}\psi_I}{\mathrm{D}s} = \frac{\mathrm{d}\psi_I}{\mathrm{d}s} + \Gamma_{\alpha\mu}^\nu \dot X^\alpha \mathrm{G}^\mu{}_\nu \psi_I \ .
\end{equation}
The operator $\mathrm{G}^\mu{}_\nu$, discussed more fully in \appref{tensor}, acts on 
tensor indices \cite{DeWitt64}. For example, we have 
\begin{subequations}
\begin{align}
	\mathrm{G}^\mu{}_\nu \psi^\rho & = \delta^\rho_\nu \psi^\mu \ ,\\
    \mathrm{G}^\mu{}_\nu \psi_\rho & = -\delta^\mu_\rho \psi_\nu  \ ,
\end{align}
\end{subequations}
for contravariant and covariant vectors. The extension of $\mathrm{G}^\mu{}_\nu$ to higher rank tensors is straightforward.  Also note that the covariant derivative of $X^\alpha$ is defined by the vector
$DX^\alpha/Ds \equiv \dot X^\alpha$. Thus, the worldline coordinates behave as spacetime scalars under covariant differentiation.

The Lagrangian \eqref{EffectiveAction} is a function over a tensor bundle. That is,  $L$ depends on the position $X(s)$ as well as tensors in the tangent space of $x = X(s)$. The variation $\delta\psi_I$ is not covariant whenever the 
base point $x$ is also varied. As a result, the 
functional derivatives $\delta S/\delta X^\alpha$ and $\delta S/\delta\psi_I$ yield the equations of motion in {non--covariant combinations.}
Here we use the results of \appref{tensor} to vary the 
action in a covariant manner. 

To begin, let us define the momentum variables
\begin{subequations}\label{EffectiveMomenta}
\begin{gather}\label{EffectiveTotalMomentum}
P_\alpha \equiv \frac{\partial L}{\partial \dot{X}^\alpha} \ ,\\\label{EffectiveTensorMomentum}
\pi^I \equiv \frac{\partial L}{\partial \mathring{\psi}_I} \ .
\end{gather}
\end{subequations}
Using the result \eqref{covvarPhi} from \appref{tensor}, the covariant variation of  the action is
\begin{align}\label{EffectiveVar1}
\begin{split}
\delta S = \int_{s_0}^{s_1} \mathrm{d} s \left( P_\alpha \Delta\dot{X}^\alpha + \frac{\partial L}{\partial \phi_A} \nabla_\alpha\phi_{A}\delta X^\alpha \right.  & \\
+ \pi^I\Delta\mathring{\psi}_I + \left. \frac{\partial L}{\partial \psi_I}\Delta \psi_I  \right) &
\end{split}
\end{align}
Here, $\Delta$ is the covariant variation defined by \cite{Steinhoff}
\begin{equation}\label{defDeltapsi}
	\Delta \equiv \delta + \Gamma_{\alpha\mu}^\nu 
    \delta X^\alpha \mathrm{G}^\mu{}_\nu  \ .
\end{equation}
Because $\phi_A$ are external sources, their  
variations are given by $\delta \phi_A = (\partial\phi_A/\partial x^\alpha)
\delta X^\alpha$. Then the covariant variations of these fields are $\Delta\phi_A = \nabla_\alpha\phi_A \delta X^\alpha$.  Also observe that the covariant variation of $X^\alpha$ is defined by the vector
$\Delta X^\alpha \equiv \delta x^\alpha$. That is, the worldline coordinates behave as spacetime scalars under covariant variation.

To bring $\delta S$ into a form that will provide the equations of motion, we must swap the order of the variations and time derivatives in $\Delta\dot{X}^\alpha$ and $\Delta \mathring{\psi}^\alpha$. The covariant variation and covariant derivative do not commute; in general, we have the following relation from Eq.~\eqref{commutatorDeltaD}:
\begin{gather}\label{commutator-DeltaDs}
\left[\Delta, \frac{\mathrm{D}}{\mathrm{D}s}\right] = R^\sigma{}_{\rho\mu\nu}\delta X^\mu \dot{X}^\nu \mathrm{G}^\rho{}_{\sigma} \ .
\end{gather}
This yields the results 
\begin{subequations}\label{commutator-velocities}
\begin{gather}
\Delta\dot{X}^\alpha = \frac{\mathrm{D}}{\mathrm{D}s}\left(\delta X^\alpha\right) \ ,\\
\Delta \mathring{\psi}_I = \frac{\mathrm{D}}{\mathrm{D}s} \left(\Delta\psi_I\right) + R^{\sigma}{}_{\rho\mu\nu}\delta X^\mu\dot{X}^\nu \mathrm{G}^\rho{}_{\sigma} \psi_I \ ,
\end{gather}
\end{subequations}
since the worldline coordinates $X^\alpha$ behave as scalar fields under covariant differentiation and covariant {variation.  

Now} integrate by parts. The endpoint terms vanish if we assume that $X^\alpha$ and $\psi_I$ are fixed at $s_0$ and $s_1$. The variation of the action becomes 
\begin{align}\label{steinhoff-var-2}
\begin{split}
\delta S =& \int_{s_0}^{s_1} \mathrm{d} s \,\left[ \left(- \mathring{P}_\alpha  + R^\sigma{}_{\rho\alpha\beta}\dot{X}^\beta \pi^I \mathrm{G}^\rho{}_{\sigma}\psi_I\phantom{\frac{\partial L}{\partial \phi_A}}\right.\right.\\
&+\left.\left.\nabla_\alpha\phi_A \frac{\partial L}{\partial \phi_A} \right)\delta X^\alpha + \left(\frac{\partial L}{\partial \psi_I} - \mathring{\pi}^I\right)\Delta \psi_I \right] \ .
\end{split}
\end{align}
The coefficients of $\delta X^\alpha$ and $\Delta\psi_I$ give the equations of motion
\begin{subequations}\label{steinhoff-eom}
\begin{gather}\label{steinhoff-eomp}
\mathring{P}_\alpha = -R_{\alpha\beta}^{\quad \rho\sigma}\dot{X}^\beta \pi^I \mathrm{G}_{\rho\sigma}\psi_I + \nabla_\alpha\phi_A\frac{\partial L}{\partial\phi_A} \ , \\\label{steinhoff-eompsi}
\mathring{\pi}^I = \frac{\partial L}{\partial \psi_I}  \ .
\end{gather}
\end{subequations}
Equation \eqref{defDeltapsi} shows that the variation $\Delta\psi_I$ is a sum of non--covariant terms proportional to 
$\delta\psi_I$ and $\delta X^\alpha$. From this we see that the equation of motion 
$\delta S/\delta\psi_I = 0$ (with $X^\alpha$ fixed) coincides with \eqref{steinhoff-eompsi}, while the equation of motion 
$\delta S/\delta X^\alpha = 0$ (with $\psi_I$ fixed) is a non--covariant
combination of the covariant Eqs.~\eqref{steinhoff-eom}

Spin is defined as
\begin{equation}\label{steinhoff-spin}
S_{\rho\sigma} \equiv 2\pi^I \mathrm{G}_{[\rho\sigma]}\psi_I \ .
\end{equation}
This puts the equation of motion \eqref{steinhoff-eomp} into the form of the MPD 
equation \eqref{SteinhoffMPDp}. 
The equation of motion for the spin variable itself is obtained from the covariant 
derivative of the definition \eqref{steinhoff-spin}, which yields $\mathring S_{\rho\sigma} = 2\mathring \pi^I \mathrm{G}_{[\rho\sigma]}\psi_I 
+ 2\pi^I \mathrm{G}_{[\rho\sigma]}\mathring\psi_I$. With the equation of motion 
(\ref{steinhoff-eompsi}) this becomes
\begin{equation}\label{Sdotdirect}
	\mathring S_{\rho\sigma} = 2\frac{\partial L}{\partial \psi_I} 
    \mathrm{G}_{[\rho\sigma]}\psi_I 
    + 2\pi^I \mathrm{G}_{[\rho\sigma]}\mathring\psi_I \ .
\end{equation}
We now make use of \eqref{ChainG}, which follows from the requirement of general 
covariance. This equation tells us that the operator $\mathrm{G}^\alpha{}_\beta$ acting on the Lagrangian $L$ of \eqref{EffectiveAction} follows
a ``chain rule". Since $L$ is a scalar, we have 
\begin{align}\label{chainruleapplied}
\begin{split}
	0 = &\frac{\partial L}{\partial\phi_A} \mathrm{G}^\alpha{}_\beta \phi_A 
    + \frac{\partial L}{\partial \dot X^\mu} \mathrm{G}^\alpha{}_\beta \dot X^\mu \\
    &+ \frac{\partial L}{\partial \psi_I} \mathrm{G}^\alpha{}_\beta \psi_I
    + \frac{\partial L}{\partial \mathring\psi_I} \mathrm{G}^\alpha{}_\beta \mathring\psi_I
\end{split}
\end{align}
We can lower the index $\alpha$ on $\mathrm{G}^\alpha{}_\beta$ and antisymmetrize. Using the 
notation \eqref{EffectiveMomenta} for the momenta, we have 
\begin{align}
\begin{split}
	0 = &\frac{\partial L}{\partial\phi_A} \mathrm{G}_{[\alpha\beta]} \phi_A 
    + P_{[\beta} \dot X_{\alpha]} \\
    &+ \frac{\partial L}{\partial \psi_I} \mathrm{G}_{[\alpha\beta]} \psi_I
    + \pi^I \mathrm{G}_{[\alpha\beta]} \mathring\psi_I
\end{split}
\end{align}
This result can be used to rewrite the time derivative of the spin from 
\eqref{Sdotdirect} as the MPD equation \eqref{SteinhoffMPDS}. 

To summarize, we have shown that the equations of motion \eqref{steinhoff-eom} that follow from 
the action \eqref{EffectiveAction}, along with the definition \eqref{steinhoff-spin} for spin, yield the { MPD} equations 
\begin{subequations}\label{SteinhoffMPD2}
\begin{gather}\label{SteinhoffMPDp2}
\mathring P_\alpha = -\frac{1}{2}R_{\alpha\beta\mu\nu}\dot{X}^\beta S^{\mu\nu} + (\nabla_\alpha\phi_A)\frac{\partial L}{\partial\phi_A} \ , \\\label{SteinhoffMPDS2}
\mathring S_{\alpha\beta} = 2P_{[\alpha}\dot{X}_{\beta]} - 2(\mathrm{G}_{[\alpha\beta]}\phi_A)\frac{\partial L}{\partial\phi_A} 
\end{gather}
\end{subequations}
in the form {(\ref{SteinhoffMPD})}.

The key difference between this result and {previous analyses} is that we do not {require a body frame or} basis vectors  $ e_a{}^\alpha$  to define the 
orientation of the body, {and we have no need for} an angular velocity 
variable $\Omega_{\alpha\beta}$. The spin,  as defined in \eqref{steinhoff-spin}, does not rely on these constructions. 

\section{Continuous bodies written as effective particles}\label{continuous} 
Let $\mathscr{M}$ be a Riemannian manifold with coordinates $x^\mu$. We consider a continuous body in the sense defined by Carter and Quintana \cite{Carter1972}, as a congruence of world lines represented by the smooth mapping $X:\mathbb{R}\times\mathscr{S}\rightarrow \mathscr{M}$, where $\mathscr{S}$ is a differentiable manifold whose points represent 
the worldlines, and the real numbers $\mathbb{R}$ label points along each world line. $\mathscr{S}$ is called the ``matter space'' and is given coordinates $\zeta^i$ \cite{Kijowski}. The coordinate on $\mathbb{R}$ is $s$. Thus, 
functions over the congruence may be written in terms of coordinates $(s,\zeta^i)$ on $\mathbb{R}\times\mathscr{S}$, or in terms of the manifold coordinates $x^\mu$. We introduce the inverse mapping $(Z^0(x),Z^i(x))$ that satisfies $s = Z^0(X(s,\zeta))$ and $\zeta^i = Z^i(X(s,\zeta))$. 

The derivatives of $X^\mu$ and $(Z^0,Z^i)$ satisfy \cite{Brown96}
\begin{subequations}\label{InverseConditions}
\begin{gather}\label{InverseConditions1}
\dot{X}^\mu Z^0_{,\nu} + X^\mu_{,i} Z^i_{,\nu} = \delta^\mu_\nu \ , \\\label{InverseConditions2}
X^\mu_{,i} Z^j_{,\mu} = \delta^j_i \ , \\\label{InverseConditions3}
\dot{X}^\mu Z^0_{,\mu} = 1 \ , \\\label{InverseConditions4}
X^\mu_{,i} Z^0_{,\mu} = 0 \ ,\\
\dot{X}^\mu Z^i_{,\mu} = 0 \ ,
\end{gather}
\end{subequations}
where $\dot{X}^\mu = \partial X^\mu/\partial s$. 

We introduce the notations
\begin{subequations}
\begin{gather}
\frac{\mathcal{D}}{\mathcal{D}s}\equiv \dot{X}^\mu\nabla_\mu = \frac{\partial}{\partial s} +\Gamma^\sigma_{\mu\rho}\dot{X}^\mu \mathrm{G}^\rho{}_\sigma \ ,\\
\frac{\mathcal{D}}{\mathcal{D}\zeta^i}\equiv X^\mu_{,i}\nabla_\mu = \frac{\partial}{\partial \zeta^i} +\Gamma^\sigma_{\mu\rho}X^\mu_{,i} \mathrm{G}^\rho{}_\sigma \ ,
\end{gather}
\end{subequations}
using the generators $\mathrm{G}^\rho{}_\sigma$ from \appref{tensor}, and following Vines' use of $\mathcal{D}$ as a covariant partial derivative \cite{Vines16}. As before, we may also abbreviate $\mathcal{D}/\mathcal{D}s$ with the circle (e.g. $\mathring{A}$) and $\mathcal{D}/\mathcal{D}\zeta^i$ with semicolons (e.g. $A_{;i}$).

We consider a fairly general action assuming that the body is subject to no external forces, though it may be subject to internal forces mediated by the first spatial derivative $X^\mu_{,i}$:
\begin{equation}\label{Continuous-action}
S[X] = \int_{s_0}^{s_1}\mathrm{d}s\int_{\mathscr{S}}\mathrm{d}^3\zeta\,\mathcal{L}\left(\zeta,g_{\mu\nu}(X),\dot{X}^\mu,X^\mu_{,i}\right) \ .
\end{equation}
We will generally restrict this action to be reparametrization invariant, although the results of this section do not depend on that assumption. When reparametrization invariance is enforced,  the above action {coincides with} DeWitt's elastic body \cite{DeWitt62,Brown16}
\begin{equation}\label{elastic}
S[X] = -\int_{s_0}^{s_1}\mathrm{d}s\int_{\mathscr{S}}\mathrm{d}^3\zeta\,\alpha\rho(\zeta,f_{ij}) \ ,
\end{equation}
where $\alpha = \sqrt{-\dot{X}^\mu \dot{X}_\mu}$ and $f_{ij}$ is the fleet metric $f_{ij} = (g_{\mu\nu}+U_\mu U_\nu)X^\mu_{,i}X^\nu_{,j}$, with $U^\mu = \dot{X}^\mu/\alpha$ being the four-velocity field.

In this section we study some basic properties of the action \eqref{Continuous-action} and show that it can be expressed in the form {of the effective particle action} \eqref{EffectiveAction}. Doing so gives definitions of $P_\alpha$ and $S_{\alpha\beta}$ identical to Dixon's definitions \eqref{DixonSnP} \cite{Dixon70-1}. 

We start by varying the action to determine the equations of motion. With the exception of its dependence on $\zeta$ (which does not affect the variation), the Lagrangian density $\mathcal{L}$ is a function of $X$ through the metric field $g_{\mu\nu}$ and is also a function of vectors $\dot{X}^\mu$ and $X^\mu_{,i}$. This means we can use the methods of \appref{tensor}, specifically Eq.~(\ref{covvarPhi}), to vary the action.  We define the canonical momentum density and stress density as 
\begin{subequations}
\begin{eqnarray}
 {\mathcal{P}}_\mu & \equiv & \frac{\partial  {\mathcal{L}}}{\partial \dot{X}^\mu} \ ,\\
 {\mathcal{S}}_\mu^{i} & \equiv & \frac{\partial  {\mathcal{L}}}{\partial X^\mu_{,i}} \ .
\end{eqnarray}
\end{subequations}
Then the variation with respect to $X^\mu$ is 
\begin{equation}\label{deltaSforcontinuum}
\delta S = \int_{s_0}^{s_1}\mathrm{d}s\int_{\mathscr{S}}\mathrm{d}^3\zeta \left[\mathcal{P}_\mu\Delta\dot{X}^\mu + \mathcal{S}_\mu^i\Delta X^\mu_{,i} \right] \ .
\end{equation}
Note that, since the metric is treated as an external field, its
covariant variation $\Delta g_{\mu\nu}(X) = \nabla_\alpha g_{\mu\nu} \delta X^\alpha$ vanishes. 

For the $\Delta \dot{X}^\mu$ term in Eq.~(\ref{deltaSforcontinuum}), integration by parts gives
\begin{align}
\begin{split}
\int_{s_0}^{s_1} \mathrm{d}s \,\mathcal{P}_\mu\Delta\dot{X}^\mu =&- \int_{s_0}^{s_1}\mathrm{d}s\frac{\mathcal{D}\mathcal{P}_\mu}{\mathcal{D}s}\delta X^\mu \\
&+ \left[\mathcal{P}_\mu\delta X^\mu\right]_{s_0}^{s_1} \ .
\end{split}
\end{align}
The last term in the above equation vanishes when we fix the initial and final configurations $X^\mu(s_0)$ and $X^\mu(s_1)$; that is, we set 
$\delta X^\mu(s_0) = \delta X^\mu(s_1) = 0$. 
We can similarly integrate by parts for the $X^\mu_{,i}$ term, with the result 
\begin{align}
\begin{split}
\int_{\mathscr{S}}\mathrm{d}^3\zeta \,\mathcal{S}_\mu^i \Delta X^\mu_{,i} = &-\int_{\mathscr{S}}\mathrm{d}^3\zeta \frac{\mathcal{D}\mathcal{S}_\mu^i}{\mathcal{D}\zeta^i}\delta X^\mu \\
&+ \int_{\partial \mathscr{S}}\mathrm{d}^2\zeta\,\eta_i \mathcal{S}_\mu^i\delta X^\mu \ .
\end{split}
\end{align}
where {$\eta_i$} is the outward-pointing vector field normal to the boundary surface $\partial \mathscr{S}$.
The variation of the action with $X^\mu$ fixed at $s_0$ and $s_1$ is then 
\begin{align}
\begin{split}\label{deltaSX}
\delta S = &\int_{s_0}^{s_1}\mathrm{d}s\int_{\mathscr{S}}\mathrm{d}^3\zeta\left[-\mathring{ {\mathcal{P}}}_\mu -  {\mathcal{S}}_{\mu;i}^i\right]\delta X^\mu\\
&+ \int_{s_0}^{s_1}\mathrm{d}s \int_{\partial\mathscr{S}}\mathrm{d}^2\zeta\, \left[ \eta_i {\mathcal{S}}^i_{\mu} \right] \delta X^\mu \ .
\end{split}
\end{align}
Since $\delta X^\mu$ is arbitrary, this gives the equations of motion and boundary conditions
\begin{subequations}\label{allfieldeqns}
\begin{align}\label{volumefieldeqn}
-\mathring{{\mathcal{P}}}_\mu -  {\mathcal{S}}_{\mu;i}^i 
= 0 \quad &(\zeta\in \mathscr{S}) \ ,\\ \label{boundaryfieldeqn}
\eta_i {\mathcal{S}}^i_{\mu} = 0 \quad &(\zeta\in\partial\mathscr{S}) \ .
\end{align}
\end{subequations}
By integrating \eqref{volumefieldeqn} over the matter space $\mathscr{S}$, and using the boundary \eqref{boundaryfieldeqn}, we obtain  a continuity equation for the total momentum within the body.

The stress--energy tensor is defined by 
\begin{equation}\label{stressenergy}
T^{\mu\nu} = \frac{2}{\sqrt{-g}}\frac{\partial {\mathcal{L}}}{\partial g_{\mu\nu}} \ .
\end{equation}
This can be rewritten in terms of the canonical momentum and stress densities as
\begin{equation}\label{ContinuousStressEnergy}
T^{\mu\nu} = \frac{1}{\sqrt{-g}}\left(\mathcal{P}^\mu\dot{X}^\nu+\mathcal{S}^{\mu i}X^\nu_{,i}\right) \ .
\end{equation}
by using the chain rule (\ref{ChainG}) for $G^\alpha{}_\beta$. 
From Eq.~\eqref{InverseConditions} we see that $\mathcal{P}^\mu = -\sqrt{h}T^{\mu\nu}n_{\nu}$ where $h$ is the determinant of the induced metric $h_{ij}=g_{\mu\nu}X^\mu_{,i}X^\nu_{,j}$ on surfaces $\Sigma(s)$ of constant $s$. Also, $n_\mu$ is the future-pointing unit vector field normal to $\Sigma(s)$. 

We now turn to the primary goal of this section: to show that the continuous body action \eqref{Continuous-action} can be written in the form of the effective particle action \eqref{EffectiveAction}. This requires mapping the information about the congruence of geodesics to the tangent space of a point on a fiducial world line. To prepare for this analysis, we first make a change of notation by placing
primes on coordinates and tensors associated with the continuous body.  In particular, we will use $X':\mathbb{R}\times\mathscr{S}\rightarrow \mathscr{M}$ rather than $X$ to denote the congruence of worldlines. For tensors associated with the body, we place the prime on the indices: for example, the  (unnormalized) velocity, momentum density and stress density are $\dot X^{\mu'}$, $\mathcal{P}_{\mu'}$ and $\mathcal{S}_{\mu'}^i$, respectively. 

The fiducial world line will be denoted $X^\mu(s)$. The fiducual worldline does not need to lie within the material body. If it does, the spacetime point $X^\mu(s)$ (for a given value of $s$) does not need to lie on the $s={\rm const}$ surface.

The introduction of the fiducual worldline adds a gauge freedom to the system, in additional to the reparametrization invariance that is already present. In later sections we will discuss how the gauge freedom can be fixed.

We make use of the bitensor and exponential map formalism of \appref{bitensor}. 
We begin by defining the exponential map from a point $X^\mu(s)$ on the fiducial worldline and a vector $\xi^\mu$ in the tangent space at $X^\mu(s)$, to the point $X^{\mu'}(s,\zeta)$ in the continuous body:
\begin{equation}\label{sub}
X'(s,\zeta) = \exp\left(X(s),\xi(s,\zeta)\right) \ .
\end{equation}
Here, $\xi^\mu(s,\zeta) = -\sigma^\mu(X(s),X'(s,\zeta))$ is the vector at $X(s)$ tangent to the geodesic $\gamma(u)$ connecting $X(s)$ and $X'(s,\zeta)$, affinely parametrized such that $\gamma(0)=X$ and $\gamma(1) = X'$.  Note that the point $X'(s,\zeta)$ defined by the exponential map depends on the geometry in a neighborhood of $X(s)$. To ensure that there is a unique geodesic connecting $X(s)$ and $X'(s,\zeta)$, the size of the body must be restricted by the curvature and its derivatives at $X(s)$. The region to which it is restricted is termed the ``normal convex neighborhood'' (NCN).

With the placement of primes on the coordinates and tensors associated with points in the continuous body, the action  \eqref{Continuous-action} becomes 
\begin{equation}\label{Continuous-action-prime}
S[X'] = \int_{s_0}^{s_1}\mathrm{d}s\int_{\mathscr{S}}\mathrm{d}^3\zeta\,\mathcal{L}\left(\zeta,g_{\mu'\nu'},\dot{X}^{\mu'},X^{\mu'}_{,i}\right) \ .
\end{equation}
Equation \eqref{sub} defines a change 
of variables in the action  from $X^{\mu'}(s,\zeta)$ to 
$X^\mu(s)$ and $\xi^\mu(s,\zeta)$. We can use Eq.~\eqref{deltaxprime} to write 
$\delta X^{\mu'}$ in terms of $\delta X^\mu$ 
and the covariant variation $\Delta \xi^\mu$. The result, 
\begin{equation}\label{deltaXprimefromAppA}
\delta X^{\mu'} = K^{\mu'}{}_\alpha \delta X^\alpha 
+ H^{\mu'}{}_\alpha \Delta \xi^\alpha \ ,
\end{equation}
makes use of the Jacobi propagators, $K^{\mu'}{}_\alpha$ and $H^{\mu'}{}_\alpha$. The Jacobi propagators are defined in terms of derivatives of the exponential map. They depend on $X^\alpha(s)$, $\xi^\alpha(s,\zeta)$ and the geometry in a neighborhood of $x=X(s)$. They are introduced in 
\appref{tensor} and further developed in \appref{bitensor}. Note that $H^{\mu'}{}_{\alpha}$ is invertible in the NCN, since there must be a one-to-one mapping between $\xi^\alpha$ and $X^{\mu'}$.

As a functional of $X^{\mu}(s)$ and $\xi^\mu(s,\zeta)$,  the variation of the action is 
\begin{widetext}
\begin{eqnarray}\label{newdeltaS}
\delta S & = & \int_{s_0}^{s_1}\mathrm{d}s\int_{\mathscr{S}}\mathrm{d}^3\zeta\,
\left[ - \mathcal{P}_{\mu'} - \mathcal{S}^i_{\mu';i}\right] \left( K^{\mu'}{}_\alpha \delta X^\alpha 
+ H^{\mu'}{}_\alpha \Delta \xi^\alpha \right) \nonumber \\
& & + \int_{s_0}^{s_1}\mathrm{d}s\int_{\partial\mathscr{S}}\mathrm{d}^2\zeta\,
\left[ \eta_i\mathcal{S}^i_{\mu'} \right]  \left( K^{\mu'}{}_\alpha \delta X^\alpha 
+ H^{\mu'}{}_\alpha \Delta \xi^\alpha \right) \ ,
\end{eqnarray}
\end{widetext}
where we have used the results from Eq.~\eqref{deltaSX}. 
Extremization of $S$ with respect to $\xi^\alpha$ yields the equations 
of motion 
\begin{subequations}\label{allfieldeqnagain}
\begin{align}\label{volumefieldeqnagain}
\left[ -\mathring{ {\mathcal{P}}}_{\mu'} -  {\mathcal{S}}_{\mu';i}^i \right] 
H^{\mu'}{}_\alpha
= 0 \quad &(\zeta\in \mathscr{S}) \ ,\\ \label{boundaryfieldeqnagain}
\left[ \eta_i {\mathcal{S}}^i_{\mu'} \right] H^{\mu'}{}_\alpha = 0 \quad &(\zeta\in\partial\mathscr{S}) \ .
\end{align}
\end{subequations}
The propagator $H^{\mu'}{}_\alpha$ is invertible in the NCN, so these equations are clearly equivalent to the equations of motion obtained by 
extremizing the action with respect to $X^{\mu'}$ (Eqs.~\eqref{allfieldeqns} with primes on the spacetime indices). 

The variation of the action  with respect to the fiducial worldline $X^\alpha$ must be handled with care, since $\delta X^\alpha$ on the boundary $\partial{\mathscr{S}}$ is not independent of 
$\delta X^\alpha$ in the bulk $\mathscr{S}$. To isolate $\delta X^\alpha$ in Eq.~\eqref{newdeltaS}, we must convert the surface term to a volume 
integral. This yields  
\begin{equation}
\label{volumefieldeqnint}
\int_{\mathscr{S}}\mathrm{d}^3\zeta \left[ -\mathring{ {\mathcal{P}}}_{\mu'} K^{\mu'}{}_{\alpha} + {\mathcal{S}}_{\mu'}^i 
K^{\mu'}{}_{\alpha;i} \right] = 0  \quad (\zeta\in \mathscr{S}) 
\end{equation}
for the equation of motion that comes from extremization of $S$ with respect to 
$X^{\mu}$. This equation is a combination of the equations  
\eqref{allfieldeqnagain}. To see this, we first multiply Eq.~\eqref{volumefieldeqnagain} by 
$\overset{-1}{H}{}^{\alpha}{}_{\nu'} K^{\nu'}{}_\beta$ and integrate over the matter space $\mathscr{S}$. We then integrate by parts on the 
term ${\mathcal{S}}_{\mu';i}^i K^{\mu'}{}_\beta$. The boundary 
term vanishes by virtue of Eq.~\eqref{boundaryfieldeqnagain} (multiplied 
by $\overset{-1}{H}{}^{\alpha}{}_{\nu'} K^{\nu'}{}_\beta$). 
The remaining volume integral is precisely the equation of motion \eqref{volumefieldeqnint}. 

The analysis above shows that the equations of motion obtained by varying $X^{\mu}$ and $\xi^\mu$ are not independent; this is a consequence of the fact that $X^{\mu}$ and $\xi^\mu$ constitute a larger set of variables than $X^{\mu'}$. The variables $X^{\mu}$ and $\xi^\mu$ contain a new gauge freedom, not present in the original variables $X^{\mu'}$, which is the freedom to choose the fiducial world line. 

{The action \eqref{Continuous-action-prime} can be written explicitly in terms of the new variables $X^\mu$ and $\xi^\mu$. To show this, we first}
use the general variation \eqref{deltaXprimefromAppA} to expand 
the derivatives of $X^{\mu'}$ as 
\begin{subequations}\label{chainruleall}
\begin{eqnarray}
\dot{X}^{\mu'} & = & K^{\mu'}{}_{\alpha} \dot{X}^\alpha + H^{\mu'}{}_{\alpha} \mathring{\xi}^\alpha \label{chainruledot}\ ,\\
X^{\mu'}_{,i} & = & H^{\mu'}{}_{\alpha}\xi^\alpha_{,i} \label{chainrulei}\ .
\end{eqnarray}
\end{subequations}
As a functional of $X^\alpha(s)$ and $\xi^\alpha(s,\zeta)$, the Lagrangian density from 
{Eq.~\eqref{Continuous-action-prime}} becomes 
\begin{equation}\label{LagDensIntermediate}
{\mathcal{L}} = {\mathcal{L}}\left( \zeta,g_{\mu'\nu'},K^{\mu'}{}_\alpha \dot X^\alpha + H^{\mu'}{}_\alpha \mathring\xi^\alpha, H^{\mu'}{}_\alpha \xi^\alpha_{,i} \right) 
\end{equation} 
where the argument of $g_{\mu'\nu'}(X')$ is written in terms 
of $X(s)$ and $\xi(s,\zeta)$ using the exponential map \eqref{sub}.
Next, we make use of the parallel propagator 
$g^\alpha{}_{\mu'}$ defined in \appref{bitensor}. 
Since the Lagrangian density ${\mathcal{L}}$ is a scalar, it depends only on scalar combinations of its arguments. Furthermore, it is a property of the propagator that contractions between contravariant and covariant indices are preserved; {\em i.e.} $A^{\mu'}g^\alpha{}_{\mu'}g^{\nu'}{}_\alpha B_{\nu'} = A^{\mu'}B_{\mu'}$. Therefore, we see that the Lagrangian density can be written as
\begin{widetext}
\begin{equation}\label{new-lagrangiandensity}
\mathcal{L} 
 = {\mathcal{L}}\left(\zeta,g_{\alpha\beta},g^\alpha{}_{\mu'}K^{\mu'}{}_\beta \dot X^\beta + g^\alpha{}_{\mu'}H^{\mu'}{}_\beta \mathring\xi^\beta, g^\alpha{}_{\mu'}H^{\mu'}{}_\beta \xi^\beta_{,i} \right) \ .
\end{equation}
The action for the {continuous body} is 
\begin{align}\label{SteinhoffFinal}
S[X,\xi] = \int_{s_0}^{s_1} \mathrm{d}s\, L 
= \int_{s_0}^{s_1} \mathrm{d}s\, \int_{\mathscr{S}}\mathrm{d}^3\zeta\, 
{\mathcal{L}}\left(\zeta,g_{\alpha\beta},g^\alpha{}_{\mu'}K^{\mu'}{}_\beta \dot X^\beta + g^\alpha{}_{\mu'}H^{\mu'}{}_\beta \mathring\xi^\beta, g^\alpha{}_{\mu'}H^{\mu'}{}_\beta \xi^\beta_{,i} \right) \ ,
\end{align}
\end{widetext}
with the Lagrangian defined by
$L = \int_{\mathscr{S}}\mathrm{d}^3\zeta\, {\mathcal{L}}$. 

The Lagrangian density \eqref{new-lagrangiandensity} is constructed entirely from 
tensor fields defined in the tangent spaces of points $X^\mu(s)$ along the fiducial world line. Hence, the action \eqref{SteinhoffFinal} describes an effective particle with worldline $X^\mu(s)$, carrying internal degrees of freedom described 
by the vectors $\xi^\mu(s)$. The effective particle interacts nonlocally with the geometry in a neighborhood of the worldline; this dependence is contained in the tensors $g^\alpha{}_{\mu'}H^{\mu'}{}_{\beta}$ and $g^\alpha{}_{\mu'}K^{\mu'}{}_{\beta}$. In Eqs.~(\ref{HK-exp}) we show the series expansions for $g^\alpha{}_{\mu'}H^{\mu'}{}_{\beta}$ and $g^\alpha{}_{\mu'}K^{\mu'}{}_{\beta}$.  Expressed in this way, 
$g^\alpha{}_{\mu'}H^{\mu'}{}_{\beta}$ and $g^\alpha{}_{\mu'}K^{\mu'}{}_{\beta}$ depend  on the Riemann tensor and its derivatives evaluated along the fiducial worldline. 

The effective particle action \eqref{SteinhoffFinal} takes the form of Eq.~\eqref{EffectiveAction} with the correspondences:
\begin{subequations}
\begin{eqnarray}
\psi_I & \leftrightarrow & \xi^\alpha(\zeta)  \ , \\
\phi_A & \leftrightarrow & \left\{ g_{\alpha\beta}, R_{\mu\nu\rho\sigma}, \nabla_\alpha R_{\mu\nu\rho\sigma}, \ldots   \right\} \ ,
\end{eqnarray}
\end{subequations}
where the dots denote higher order derivatives of the Riemann tensor. 
Note that the index $I$ now represents the continuous labels $\zeta^i$, as well as the discrete spacetime index $\alpha$. This requires us to replace certain partial derivatives with functional derivatives; for example, $\partial L/\partial \psi_I$ becomes $\delta L/\delta \xi^\alpha(\zeta)$. Furthermore, a repeated index $I$ must include an integral over the matter space $\mathscr{S}$. 

The momenta as defined in Eqs.~\eqref{EffectiveMomenta} are 
\begin{subequations}
\begin{eqnarray}
P_\alpha & \equiv & \frac{\partial L}{\partial \dot{X}^\alpha} = \int_{\mathscr{S}}\mathrm{d}^3\zeta\, K^{\mu'}{}_{\alpha} {\mathcal{P}}_{\mu'} \ ,\label{dewitt-steinhoff-p}\\
\pi_\alpha(\zeta) & \equiv & \frac{\delta L}{\delta \mathring{\xi}^\alpha(\zeta)} = H^{\mu'}{}_{\alpha}  {\mathcal{P}}_{\mu'} \ .\label{pidefforcontinuous}
\end{eqnarray}
\end{subequations}
These results are most easily derived with the Lagrangian density written in the form of Eq.~\eqref{LagDensIntermediate}. 
We also note the result 
\begin{equation}
\frac{\delta L}{\delta \xi^\alpha_{,i}} = H^{\mu'}{}_{\alpha} 
\mathcal{S}^i_{\mu'}
\end{equation}
satisfied by the momentum density. 
From the definition \eqref{steinhoff-spin}, the spin {is}
\begin{equation}\label{Sdefforcontinuous}
S_{\alpha\beta} = 2\int_{\mathscr{S}}\mathrm{d}^3\zeta\ \pi_\mu \mathrm{G}_{[\alpha\beta]} \xi^\mu = 2\int_{\mathscr{S}}\mathrm{d}^3\zeta\ \xi_{[\alpha}\pi_{\beta]} \ .
\end{equation}
Using $\mathrm{d}\Sigma_{\nu'} = -n_{\nu'} \sqrt{h'}\,d^3\zeta$, our earlier result that $\mathcal{P}_{\mu'} = -\sqrt{h'}n_{\nu'}T^{\nu'}_{\mu'}$, and the definition $\xi^\alpha = -\sigma^\alpha$, we see that the above definitions for momentum 
and spin coincide with Dixon's definitions in Eqs.~\eqref{DixonSnP}.

The equations of motion for the effective particle, as derived in \secref{mpd}, must be generalized to account for the fact that the index $I$ now includes the continuous labels $\zeta^i$. 
The variation of the action, from Eq.~\eqref{EffectiveVar1}, {becomes}
\begin{widetext}
\begin{equation}
\delta S = \int_{s_0}^{s_1} \mathrm{d}s \left( P_\alpha \Delta \dot X^\alpha + \frac{\partial L}{\partial R_{\mu\nu\rho\sigma}} \nabla_\alpha R_{\mu\nu\rho\sigma} \delta X^\alpha + \cdots
+ \int_{\mathscr{S}}\mathrm{d}^3\zeta\, 
\left[ \pi_\alpha\Delta\mathring\xi^\alpha + 
\frac{\delta L}{\delta \xi^\alpha} \Delta\xi^\alpha 
+ \frac{\delta L}{\delta \xi^\alpha_{,i}} \Delta \xi^\alpha_{,i} \right] \right) \ .
\end{equation}
After integration by parts, we obtain the generalization of Eq.~\eqref{steinhoff-var-2}: 
\begin{eqnarray}
\delta S & = & \int_{s_0}^{s_1} \mathrm{d}s \left( -\mathring P_\alpha + \frac{\partial L}{\partial R_{\mu\nu\rho\sigma}} \nabla_\alpha R_{\mu\nu\rho\sigma} + \cdots +   R^\mu{}_{\nu\alpha\sigma} \dot X^\sigma \int_{\mathscr{S}}\mathrm{d}^3\zeta\, \pi_\mu \xi^\nu 
\right)\delta X^\alpha  \nonumber\\
& & + \int_{s_0}^{s_1} \mathrm{d}s \int_{\mathscr{S}}\mathrm{d}^3\zeta 
\left( -\mathring\pi_\alpha + \frac{\delta L}{\delta \xi^\alpha} - \frac{\mathcal{D}}{\mathcal{D}\zeta^i}\left( \frac{\delta L}{\delta 
\xi^\alpha_{,i}} \right) \right)  \Delta\xi^\alpha  
+ \int_{s_0}^{s_1} \mathrm{d}s \int_{\partial \mathscr{S}}\mathrm{d}^2\zeta \left(\eta_i \frac{\delta L}{\delta \xi^\alpha_{,i}} \right)  \Delta \xi^\alpha \ .
\end{eqnarray}
\end{widetext}
The equations of motion obtained by extremizing $S$ with respect to $\xi^\alpha(\zeta)$ are 
\begin{subequations}\label{eomsforpialpha}
\begin{align}
\mathring\pi_\alpha  =  \frac{\delta L}{\delta \xi^\alpha} - \frac{\mathcal{D}}{\mathcal{D}\zeta^i}\left( \frac{\delta L}{\delta 
\xi^\alpha_{,i}} \right) \quad &(\zeta \in \mathscr{S}) \ , \\
\eta_i \frac{\delta L}{\delta \xi^\alpha_{i,}} = 0 \quad &(\zeta \in \partial\mathscr{S}) \ .
\end{align}
\end{subequations}
These equations generalize the effective particle Eq.~\eqref{steinhoff-eompsi}. 
They are equivalent to Eqs.~\eqref{allfieldeqnagain}. 

The equation of motion that comes from extremization {of $S$} with respect to the fiducial particle worldline $X^\alpha$ is 
\begin{equation}\label{eomforXalpha}
\mathring{P}_\alpha  =  - R_{\alpha\beta}{}^{\rho\sigma} \dot X^\beta \int_{\mathscr{S}}\mathrm{d}^3\zeta\, \xi_\rho \pi_\sigma + \frac{\partial L}{\partial R_{\mu\nu\rho\sigma}} \nabla_\alpha R_{\mu\nu\rho\sigma}  + \cdots  \ .
\end{equation}
This equation generalizes the effective 
particle Eq.~\eqref{steinhoff-eomp}.
It is equivalent to Eq.~\eqref{volumefieldeqnint}.  As shown before, Eq.~\eqref{eomforXalpha} is redundant; it can be derived from Eqs.~\eqref{eomsforpialpha}. 

We have shown that the continuous body, described by the action \eqref{Continuous-action-prime}, can be interpreted as an effective single particle with action \eqref{SteinhoffFinal}. This result
still holds when the Lagrangian density depends on higher-order spatial derivatives of $X^{\mu'}$. In that case, the formulae for converting derivatives of $X^{\mu'}$ into derivatives of $\xi^\alpha$ and bitensors are more complicated. Nevertheless, it is still true that the final Lagrangian density
can be written in terms of a  $X^\mu(s)$,  $\xi^\alpha(s,\zeta)$ and their derivatives, in combinations that depend on the geometry in a neighborhood of the fiducal worldline. In terms of a series expansion, the dependence on 
geometry appears as the Riemann tensor and its derivatives evaluated at $X^\mu(s)$. 

We can also generalize the action \eqref{Continuous-action-prime} by allowing for further dependence on $X'$ through some exterior fields. This  would not change the main result---we simply use the parallel propagator to express these fields in terms of tensors defined along the fiducial worldline.

\section{Gauge conditions}\label{conditions}
In this section we assume the action \eqref{Continuous-action}
for the continuous body is reparametrization invariant (RP-invariant). 
A reparametrization $\tilde{s}=F(s,\zeta)$ consists in replacing the coordinates $X^{\mu'}$ with $\tilde{X}^{\mu'}$ such that $\tilde{X}^{\mu'}(\tilde{s},\zeta) = X^{\mu'}(s,\zeta)$.
In this case, the action can be written in the form \eqref{elastic} for an 
elastic body \cite{Brown16}.

A natural way to remove the gauge freedom for the continuous body is to choose the ``proper time gauge" $\dot X^{\mu'} \dot X_{\mu'} = -1$. With this condition, 
the separation $ds$ between neighboring constant $s$ surfaces coincides 
with the proper time interval measured along each of the worldlines.

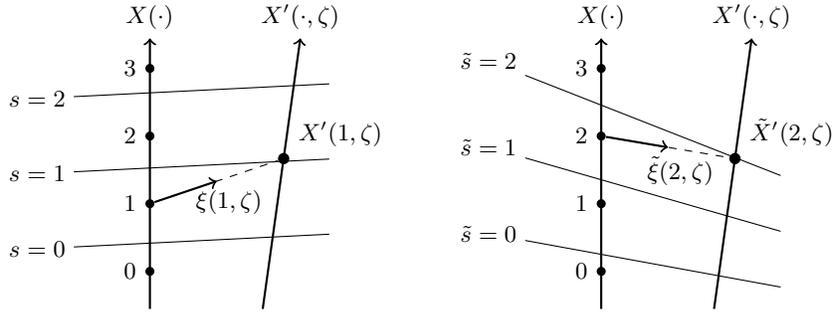
\begin{figure*}
\centering
\begin{tikzpicture}
\node (node1a) at (0,-0.2) {$s=0$};
\node (node1b) at (4,0) {};
\node (node2a) at (0,0.8) {$s=1$};
\node (node2b) at (4,1) {};
\node (node3a) at (0,1.8) {$s=2$};
\node (node3b) at (4,2) {};
\draw (node1a) -- (node1b);
\draw (node2a) -- (node2b);
\draw (node3a) -- (node3b); 
\draw [->, thick] (1.5,-1) -- (1.5,2.6) node [above] {$X(\cdot)$};
\draw [->, thick] (3,-1) -- (3.5,2.6) node [above] {$X'(\cdot,\zeta)$};
\node (dotzero) at (1.5,-0.5) [circle, fill, inner sep=0pt, minimum size=0.12cm,label=left:$0$] {};
\node (dotone) at (1.5,0.4) [circle, fill, inner sep=0pt, minimum size=0.12cm,label=left:$1$] {};
\node (dottwo) at (1.5,1.3) [circle, fill, inner sep=0pt, minimum size=0.12cm,label=left:$2$] {};
\node (dotthree) at (1.5,2.2) [circle, fill, inner sep=0pt, minimum size=0.12cm,label=left:$3$] {};
\node (atX) at (3.278,1) [circle, fill, inner sep=0pt, minimum size=0.15cm,label={[label distance=0.4]30:$X'(1,\zeta)$}] {};
\draw [dashed] (dotone) -- (atX);
\draw [->, thick] (dotone) -- ($(dotone)!0.5!(atX)$) node [below] {$\quad \xi(1,\zeta)$};
\begin{scope}[shift={(6,0)}]
\node (node1a) at (0,0) {$\tilde s=0$};
\node (node1b) at (4,-0.73) {};
\node (node2a) at (0,1.15) {$\tilde s=1$};
\node (node2b) at (4,0) {};
\node (node3a) at (0,2.3) {$\tilde s=2$};
\node (node3b) at (4,0.73) {};
\draw (node1a) -- (node1b);
\draw (node2a) -- (node2b);
\draw (node3a) -- (node3b); 
\draw [->, thick] (1.5,-1) -- (1.5,2.6) node [above] {$X(\cdot)$};
\draw [->, thick] (3,-1) -- (3.5,2.6) node [above] {$X'(\cdot,\zeta)$};
\node (dotzero) at (1.5,-0.5) [circle, fill, inner sep=0pt, minimum size=0.12cm,label=left:$0$] {};
\node (dotone) at (1.5,0.4) [circle, fill, inner sep=0pt, minimum size=0.12cm,label=left:$1$] {};
\node (dottwo) at (1.5,1.3) [circle, fill, inner sep=0pt, minimum size=0.12cm,label=left:$2$] {};
\node (dotthree) at (1.5,2.2) [circle, fill, inner sep=0pt, minimum size=0.12cm,label=left:$3$] {};
\node (atX) at (3.278,1) [circle, fill, inner sep=0pt, minimum size=0.15cm,label={[label distance=0.4]30:$\tilde X'(2,\zeta)$}] {};
\draw [dashed] (dottwo) -- (atX);
\draw [->, thick] (dottwo) -- ($(dottwo)!0.5!(atX)$) node [below] {$\quad \tilde\xi(2,\zeta)$};
\end{scope}
\end{tikzpicture}
\caption{Each figure shows the fiducial worldline $X(s)$ and the 
 worldline for a generic particle $X'(s,\zeta)$ in the body. The dots on the fiducial worldline indicate 
the parameter values for $X(s)$. In the left figure, the particle worldlines are parametrized by $s$; in the right figure, the particle worldlines are parametrized by $\tilde s$. The spacetime point 
$X'(1,\zeta) = \exp(X(1),\xi(1,\zeta))$ in the left figure coincides with the spacetime point $\tilde X'(2,\zeta) = \exp(X(2),\tilde\xi(2,\zeta))$ in the right figure. The dashed line in each figure is the geodesic that defines the exponential map.}
\label{RPfigure}
\end{figure*}

The effective particle action \eqref{SteinhoffFinal} inherits RP--invariance from 
the continuous body \eqref{Continuous-action}. Recall the change of variables from $X^{\mu'}$ to $X^\mu$ and $\xi^\mu$, defined by the 
exponential map, Eq.~\eqref{sub}. RP--invariance 
consists in replacing the vector $\xi^\alpha(s,\zeta)$ with  $\tilde{\xi}^\alpha(\tilde{s},\zeta)$ such that
\begin{equation}\label{RP-eff}
\tilde X'(\tilde s,\zeta) = \exp\left(X(\tilde{s}),\tilde{\xi}(\tilde{s},\zeta)\right) \ .
\end{equation}
Note that the reparametrization $\tilde s = F(s,\zeta)$ only changes the parametrization of the particle worldlines. It does not affect the parametrization of the fiducial world line. See Figure~\ref{RPfigure}.

The fiducial worldline can be chosen arbitrarily. This freedom 
appears as a gauge symmetry for the effective particle action \eqref{SteinhoffFinal}, in addition to the RP--invariance. 
This gauge symmetry can be identified by varying the action, 
\begin{equation}
\delta S = \int \mathrm{d}s\int_{\mathscr{S}}\mathrm{d}^3\zeta\, 
\left(\frac{\delta S}{\delta X^{\mu'}}\right) \delta X^{\mu'}  \ ,
\end{equation}
where $\delta X^{\mu'}$ depends on $\delta X^\alpha$ and $\delta \xi^\alpha$ through the exponential map \eqref{sub}. Using the results of 
\appref{bitensor}, 
we find $\delta X^{\mu'} = K^{\mu'}{}_\alpha \delta X^\alpha + H^{\mu'}{}_\alpha \Delta \xi^\alpha$. This shows that the action is invariant, $\delta S = 0$, for any variation satisfying
\begin{equation}\label{WL-eff}
\Delta \xi^\beta = \sigma^\beta{}_{\mu'} K^{\mu'}{}_\alpha 
\delta X^\alpha \ .
\end{equation}
This invariance holds independent of the equations of motion. 

For the effective particle, the reparametrization invariance \eqref{RP-eff} and {fiducial} worldline invariance \eqref{WL-eff} are independent symmetries, each requiring their own conditions for gauge fixing. RP--invariance { is characterized} by 1 real function of 4 real parameters. Worldline invariance is characterized by 4 real functions of 1 real parameter. 

{There are many ways to fix RP-invariance for the effective particle. We will consider two different 
conditions, namely, the ``proper time" gauge and the ``normal" gauge. For the 
fiducial worldline invariance, it is generally convenient to impose $\dot X^\mu \dot X_\mu  = -1$. This partially fixes the gauge by setting the fiducial worldline parameter equal to proper time. The remaining freedom in the fiducial worldline is removed by imposing a center of mass condition. We show that for a specific definition of the center of mass, the center of mass condition is related to a spin supplementary condition.}

{The} proper time gauge for an effective particle is an approximation to the conditions $\dot X^{\mu'}\dot X_{\mu'}= -1$, where $X^{\mu'}$ is defined in terms of $X^\alpha$ and $\xi^\alpha$ by the exponential map. In our examples in the next sections, we will take this approximation to third order in $\epsilon$, where we suppose  $|\xi|/\ell < \epsilon$ for any relevant length scale
{$\ell$. For example, $\ell$ would include the radius of curvature $\sim 1/\sqrt{|R|}$. This condition allows for convergence of series expansions in  $\xi$. We also impose $|\mathring{\xi}|/\alpha<\epsilon$. This second condition  requires the relative motions of the particles to be much less than the speed of light (so that vibrational energy is not comparable to the total mass--energy).

Using} Eqs.~\eqref{chainruleall} and \eqref{HK-exp}, we have 
\begin{widetext}
\begin{align}
\begin{split}
g^{\alpha}{}_{\mu'}\dot{X}^{\mu'} &= \left(\delta^\alpha_\beta - \frac{1}{2}R^\alpha{}_{\xi\beta\xi}-\frac{1}{6} R^\alpha{}_{\xi\beta\xi;\xi}\right)\dot{X}^\beta + \left(\delta^\alpha_\beta - \frac{1}{6}R^\alpha{}_{\xi\beta\xi}\right)\mathring{\xi}^\alpha + \mathcal{O}(\epsilon^4)\\
&= \dot{X}^\alpha + \mathring{\xi}^\alpha - \frac{1}{2}R^\alpha{}_{\xi\dot{X}\xi} - \frac{1}{6}R^\alpha{}_{\xi\mathring{\xi}\xi} - \frac{1}{6}R^\alpha{}_{\xi\dot{X}\xi;\xi} + \mathcal{O}(\epsilon^4) \ ,
\end{split}
\end{align}
which gives
\begin{align}\label{Xdotnorm}
\begin{split}
\dot X^{\mu'}\dot X_{\mu'} = &\dot X_\alpha \dot X^\alpha + 2\dot{X}_\alpha \mathring{\xi}^\alpha + \mathring{\xi}^2 - R_{\dot{X}\xi\dot{X}\xi} - \frac{4}{3}R_{\dot{X}\xi\mathring{\xi}\xi} - \frac{1}{3}R_{\dot{X}\xi\dot{X}\xi;\xi}+ \mathcal{O}(\epsilon^4) \ .
\end{split}
\end{align}
\end{widetext}
The occurrence of $\dot{X}$ and $\xi$ as indices on the Riemann curvature tensor indicates contraction of those indices with the specified vectors. We now partially fix the fiducial worldline by setting $\dot X_\alpha \dot X^\alpha = -1$. Then the proper time gauge condition  $\dot X^{\mu'}\dot X_{\mu'} = -1$ yields
\begin{equation}\label{PT-gauge}
2\dot{X}_\alpha \mathring{\xi}^\alpha + \mathring{\xi}^2 - R_{\dot{X}\xi\dot{X}\xi} - \frac{4}{3}R_{\dot{X}\xi\mathring{\xi}\xi} - \frac{1}{3} R_{\dot{X}\xi\dot{X}\xi;\xi} + \mathcal{O}(\epsilon^4) = 0 \ ,
\end{equation}
to third order in $\epsilon$. 

An alternative to the proper time gauge is the ``normal gauge", in which we simply suppose the vectors $\xi^\alpha(s,\zeta)$ inhabit the subspace normal to some timelike vector field $V^\alpha(s)$ defined along the world line, that is,
\begin{equation}\label{Normal}
V_\alpha(s) \xi^\alpha(s,\zeta) = 0 \ .
\end{equation}
The vector field can be a defined in various ways; for example, $V^\alpha$ 
might equal four--velocity $U^\alpha$ or the momentum $P^\alpha$.

{With $\dot X^\alpha \dot X_\alpha = -1$, three degrees of freedom remain to be fixed for the fiducial worldline.} Two options are a ``center of mass'' condition {and} {or} a ``spin supplementary'' condition $V_\alpha S^{\alpha\beta}=0$. We will consider a natural formulation of the center of mass condition which is equivalent to a spin supplementary condition up to quadrupole order, though further multipole corrections exist.

The center of mass at parameter time $s$ can be defined as the Frechet-Karcher mean of the surface $\Sigma(s)$. Specifically, $X(s)$ is the center of mass of $\Sigma(s)$ if it minimizes the function \cite{Karcher}
\begin{equation}
f(y) \equiv \int_{\mathscr{S}} \mathrm{d}^3\zeta\,w(\zeta)\, \sigma(y,X'(s,\zeta))  \ ,
\end{equation}
where $w(\zeta)$ can be any density defined on $\mathscr{S}$. Essentially, this definition tells us that the center of mass minimizes a weighted average of the squared distance between itself and every other point on the surface. It may be the case that the point $x = X(s)$ that minimizes $f(y)$ does not lie on the surface $\Sigma(s)$.

The most useful form of the center of mass condition is found by setting $\partial_\alpha f = 0$ at the minimum. With the relation \eqref{syngetoexp} this yields $\Xi_\alpha = 0$, where 
\begin{equation}\label{COM}
\Xi_\alpha \equiv \int_{\mathscr{S}} d^3\zeta\, w\, \xi_\alpha
\end{equation}
defines the center of mass

One natural choice for the density weight is $w = \sqrt{h'}n^{\mu'}n^{\nu'}T_{\mu'\nu'}$, where $n^{\mu'}$ is the unit vector field normal to the surface $\Sigma(s)$ of constant $s$ and $h'$ is the 
determinant of the induced metric on $\Sigma(s)$. This $w$ is the energy density relative to a fleet of observers at rest in $\Sigma(s)$. If we choose the normal gauge, then $\Sigma(s)$ is (a subset of) the surface of geodesics passing through $X(s)$ which are normal to $V_\alpha$ at $X(s)$.  This gauge choice has been considered by Costa and Natario \cite{Costa15}, but using different definitions of $P_\alpha$ and $S_{\alpha\beta}$ than those used in this paper. They showed that with the normal gauge (with their definitions of spin and momentum), the center of mass condition $\Xi_\alpha = 0$ with 
$ w= -\sqrt{h'}Z^{0,\mu'}n^{\nu'}T_{\mu'\nu'}$ is exactly equivalent to a spin supplementary condition $V_\alpha S^{\alpha\beta}=0$. We produce a similar result for our definitions of $P_{\alpha}$, $S_{\alpha\beta}$ and $w$.

The quickest way to derive the result is by using the series expansion of $\overset{-1}{H}{}^\alpha{}_{\mu'}n^{\mu'}$. If we differentiate the normal gauge condition \eqref{Normal} with respect to $\zeta^i$, we see that it satisfies the property $V_\alpha \xi^\alpha_{,i} = 0$. We also have $n_{\mu'}X^{\mu'}_{,i}=0$ everywhere on $\Sigma(s)$. Using the identity $\overset{-1}{H}{}^\alpha{}_{\mu'}H^{\nu'}{}_{\alpha}=\delta^{\mu'}_{\nu'}$ and the result \eqref{chainrulei}, we can write  $n_{\nu'} \delta^{\nu'}_{\mu'}X^{\mu'}_{,i}=0$ as
\begin{equation}
\left(H^{\nu'}{}_{\alpha}n_{\nu'}\right)\left(\overset{-1}{H}{}^\alpha{}_{\mu'}X^{\mu'}_{,i}\right)=H^{\nu'}{}_{\alpha}n_{\nu'} \xi^{\alpha}_{,i}=0 \ .
\end{equation}
This means that $H^{\mu'}{}_\alpha n_{\mu'}$ is parallel to $V_\alpha$. Next,  set $H^{\mu'}{}_\alpha n_{\mu'} = AV_\alpha$ and write
\begin{equation}
g^{\mu'\nu'}n_{\mu'}n_{\nu'} =  A^2\left(\overset{-1}{H}{}^\alpha{}_{\mu'}\overset{-1}{H}{}^\beta{}_{\nu'}g^{\mu'\nu'}\right)V_\alpha V_\beta \ .
\end{equation}
Using the series expansion \eqref{g-exp} and $n^{\mu'}n_{\mu'} = -1$ we have
\begin{widetext}
\begin{align}
\begin{split}
A &= \left(1 -\frac{1}{3} R_{V\xi V\xi} - \frac{1}{6} \nabla_\xi R_{V\xi V\xi} + \mathcal{O}(\epsilon^4)\right)^{-\frac{1}{2}} = 1 + \frac{1}{6}R_{V\xi V\xi}+ \frac{1}{12}\nabla_\xi R_{V\xi V\xi} + \mathcal{O}(\epsilon^4) \ .
\end{split}
\end{align}
Now we can calculate
\begin{equation}
\overset{-1}{H}{}^\alpha{}_{\mu'}n^{\mu'}  
= \overset{-1}{H}{}^\alpha{}_{\mu'} g^{\mu'\nu'}  n_{\nu'} 
= \overset{-1}{H}{}^\alpha{}_{\mu'} g^{\mu'\nu'} \left(  \overset{-1}{H}{}^\beta{}_{\nu'} H^{\lambda'}{}_{\beta} \right) n_{\lambda'}  \ .
\end{equation}
Using the results above for $H^{\lambda'}{}_\beta n_{\lambda'}$ and the 
expansion \eqref{g-expHinvHinv} from \appref{bitensor}, we have
\begin{align}\label{n-exp}
\begin{split}
\overset{-1}{H}{}^\alpha{}_{\mu'}n^{\mu'} &=  \left(g^{\alpha\beta} + \frac{1}{3}R^\alpha{}_\xi{}^\beta{}_\xi+ \frac{1}{6}\nabla_\xi R^\alpha{}_\xi{}^\beta{}_\xi + \mathcal{O}(\epsilon^4)\right) \left(1 + \frac{1}{6}R_{V\xi V\xi}+ \frac{1}{12}\nabla_\xi R_{V\xi V\xi} + \mathcal{O}(\epsilon^4)\right)V_\beta\\
&= V^\alpha + \frac{1}{3}\psi^{\alpha\beta}\left(R_{\beta\xi V\xi}  + \frac{1}{2} \nabla_\xi R_{\beta\xi V\xi}\right) +\mathcal{O}(\epsilon^4) \ .
\end{split}
\end{align}
where $\psi^{\alpha\beta} \equiv g^{\alpha\beta} + \frac{1}{2}V^\alpha V^\beta$.

Now set $w = \sqrt{h'}n^{\mu'}n^{\nu'}T_{\mu'\nu'}=-n^{\mu'}\mathcal{P}_{\mu'}$, where the second equality follows from the results of \secref{continuous}, to obtain
\begin{align}
\begin{split}
\int_{\mathscr{S}}&\mathrm{d}^3\zeta\,w \,\xi_\alpha  = -\int_{\mathscr{S}}\mathrm{d}^3\zeta\,
n^{\mu'} \overset{-1}{H}{}^\beta{}_{\mu'} H^{\nu'}{}_\beta \mathcal{P}_{\nu'}\xi_\alpha 
= -2\int_{\mathscr{S}}\mathrm{d}^3\zeta\,\overset{-1}{H}{}^\beta{}_{\mu'}n^{\mu'}\xi_{[\alpha} H^{\nu'}{}_{\beta]}\mathcal{P}_{\nu'} \ .
\end{split}
\end{align}
Here we utilize the result $\overset{-1}{H}{}^{\beta}{}_{\mu'}n^{\mu'}\xi_\beta = n^{\mu'}\sigma^\beta_{\mu'}\sigma_\beta = n^{\mu'}\sigma_{\mu'} = 0$, derived from the relations in \appref{bitensor} and the fact that $\sigma_{\mu'}$ is tangent to $\Sigma(s)$ and therefore orthogonal to $n^{\mu'}$. Now replacing $\overset{-1}{H}{}^{\beta}{}_{\mu'}n^{\mu'}$ with the expansion in Eq.~\eqref{n-exp} and using the 
definition of spin from Eqs.~\eqref{pidefforcontinuous} and \eqref{Sdefforcontinuous}, we find that the center of mass satisfies 
\begin{align}\label{SSC-correction}
\begin{split}
\Xi_\alpha \equiv 
\int_{\mathscr{S}}&\mathrm{d}^3\zeta\,w\, \xi_\alpha  = V^\beta S_{\beta\alpha} - \frac{2}{3}\psi^{\beta\gamma}V^\delta R_{\gamma\rho\delta\sigma}O^{\rho\sigma}{}_{\alpha\beta}+\mathcal{O}(\epsilon^4) \ .
\end{split}
\end{align}
\end{widetext}
Here, 
\begin{equation}\label{octupolemoment}
O^{\rho\sigma}{}_{\alpha\beta} \equiv \int_{\mathscr{S}}\mathrm{d}^3\zeta\,\xi^\rho\xi^\sigma\xi_{[\alpha} H^{\mu'}{}_{\beta]}\mathcal{P}_{\mu'}
\end{equation}
is an octupole moment in the sense that it is an integral over a density field with three position vectors. It is not clear to us how this term relates to the gravitational or reduced multipoles. In any case, {Eq.~\eqref{SSC-correction} shows} that the center of mass condition {$\Xi_\alpha = 0$} with appropriately chosen weight $w$ is equivalent to a spin supplementary condition up to second order in $\epsilon$, with {an} octupole correction at third order.

\section{Example: Dust}\label{example}
An extremely simple example of a continuous body is a cloud of dust. 
This is a special case of an elastic body Eq.~\eqref{elastic} with 
the Lagrangian density $\mathcal{L} = -\alpha \rho$, where $ {\rho}(\zeta,f_{ij}) =  {\rho}(\zeta)$ is independent of the fleet metric $f_{ij}$. Thus, for the dust cloud, there is no interaction energy between particles. 

With the definition $\alpha = \sqrt{-\dot X^{\mu'}\dot X_{\mu'}}$, the dust cloud action is
\begin{equation}\label{dustcloudaction}
S[X'] = -\int_{s_0}^{s_1}\mathrm{d}s\int_{\mathscr{S}}\mathrm{d}^3\zeta\,
\rho \sqrt{-\dot X^{\mu'}\dot X_{\mu'}}  \ .
\end{equation}
The equations of motion are given in \eqref{allfieldeqns}, with 
primes placed on the spacetime indices. In this case the stress density 
$\mathcal{S}^i_{\mu'} \equiv \partial\mathcal{L}/\partial X^{\mu'}_{,i}$ 
vanishes, so the equations of motion reduce to $\mathring{\mathcal{P}}_{\mu'} = 0$. The momentum density is $\mathcal{P}_{\mu'} \equiv \partial\mathcal{L}/\partial \dot X^{\mu'} = \rho U_{\mu'}$, where 
$U_{\mu'}(s,\zeta) \equiv \dot X_{\mu'}/\sqrt{-\dot X^{\mu'}\dot X_{\mu'}}$ is the four--velocity of the particle with label $\zeta^i$. Therefore the equations of motion for the dust cloud imply $\mathring{U}_{\mu'} = 0$; as expected, each dust particle follows a geodesic. 

We now use the exponential map to write the action in terms of a fiducial 
worldline $X^{\mu}$ and vectors $\xi^\mu$. From Eq.~\eqref{Xdotnorm} we have 
\begin{widetext}
\begin{equation}
\sqrt{-\dot X^{\mu'} \dot X_{\mu'}} = \sqrt{-\dot X^2}
\left\{ 1 - U\cdot \overset{*}{\xi} - \frac{1}{2} \left(1 + U\cdot \overset{*}{\xi} \right) \left(f_{\alpha\beta} \overset{*}{\xi}{}^\alpha \overset{*}{\xi}{}^\beta - R_{U\xi U\xi} \right) + \frac{2}{3} R_{U\xi \overset{*}{\xi} \xi} + \frac{1}{6} R_{U\xi U\xi ;\xi}  + \mathcal{O}(\epsilon^4) \right\} 
\label{alphaexpanded}
\end{equation}
\end{widetext}
where we have used the abbreviations $U^\alpha \equiv \dot X^\alpha/\sqrt{-\dot X^2}$ and 
$\overset{*}{\xi}{}^\alpha \equiv \mathring\xi^\alpha/ \sqrt{-\dot X^2}$. 
The action is given to third order in $\epsilon$ by inserting 
\eqref{alphaexpanded} into \eqref{dustcloudaction}. 

The canonical momenta conjugate to $X^\alpha$ and $\xi^\alpha$  are
\begin{eqnarray}\label{Pfordust}
P_\alpha & = & \int_{\mathscr{S}}\mathrm{d}^3\zeta\,\rho 
\left\{ 
U_\alpha + \frac{1}{2} \left( f_{\beta\sigma} \overset{*}{\xi}{}^\beta \overset{*}{\xi}{}^\sigma - R_{U\xi U\xi} \right) \right. \nonumber \\
& & + \left. \left( 1 + U\cdot\overset{*}{\xi} \right) f_{\alpha\beta}\overset{*}{\xi}{}^\beta - R_{\alpha\xi U \xi} +
\mathcal{O}(\epsilon^3)
\right\} \quad
\end{eqnarray}
and
\begin{eqnarray}\label{pifordust}
\pi_\alpha  & = &  \rho\left\{ U_\alpha + \frac{1}{2} U_\alpha\left( f_{\beta\sigma}\overset{*}\xi{}^\beta \overset{*}\xi{}^\sigma - R_{U\xi U\xi} \right)  \right. \nonumber\\
& & + \left. \left( 1 + U\cdot \overset{*}{\xi}\right) f_{\alpha\beta} \overset{*}{\xi}{}^\beta 
- \frac{2}{3} R_{U\xi\alpha \xi}  + \mathcal{O}(\epsilon^3) \right\} 
\ , \qquad
\end{eqnarray}
respectively. The equations of motion \eqref{steinhoff-eom} are
\begin{eqnarray}\label{Pdoteqnfordust}
\mathring{P}_\alpha & = & \int_{\mathscr{S}}\mathrm{d}^3\zeta\,\rho \sqrt{-\dot X^2} \left\{ \left( 1 + U\cdot \overset{*}{\xi} \right) R_{\alpha UU\xi} \right. \nonumber\\
& & \qquad + \left. R_{\alpha U \overset{*}{\xi} \xi} - \frac{1}{2} R_{U\xi U\xi;\alpha} +\mathcal{O}(\epsilon^3) 
\right\} 
\end{eqnarray}
and
\begin{eqnarray}\label{pidoteqnfordust}
\mathring{\pi}_\alpha & = & \rho \sqrt{-\dot X^2} \left\{\left( 1 + U\cdot\overset{*}{\xi}\right) R_{\alpha UU\xi} + \frac{4}{3} R_{\alpha (U\overset{*}{\xi})\xi} \right. \nonumber\\
& & \quad +\left.\frac{1}{3} R_{\alpha UU\xi;\xi} - \frac{1}{6}R_{U\xi U\xi;\alpha} +\mathcal{O}(\epsilon^3) \right\}  \ .
\end{eqnarray}
Equation \eqref{Pdoteqnfordust} follows from Eq.~\eqref{pidoteqnfordust}. 
To show this, we multiply Eq.~\eqref{pidoteqnfordust} by 
$\overset{-1}{H}{}^\alpha{}_{\mu'} K^{\mu'}{}_\beta$ and integrate over the matter space $\mathscr{S}$. We can make use of the 
relations 
\begin{equation}\label{sigmaseries}
\overset{-1}{H}{}^\alpha{}_{\mu'} K^{\mu'}{}_\beta = \sigma^\alpha{}_\beta = \delta^\alpha_\beta - \frac{1}{3} R^\alpha{}_{\xi\beta\xi} + \mathcal{O}(\epsilon^3) 
\end{equation}
which follow from Eqs.~\eqref{DixonHandK} and \eqref{HK-exp} of \appref{bitensor}. The calculation leading from \eqref{pidoteqnfordust} to \eqref{Pdoteqnfordust} uses the first and second Bianchi identities. 

The spin for the dust cloud, from the definition \eqref{steinhoff-spin}, is 
\begin{widetext}
\begin{equation}\label{Spinfordust}
S_{\alpha\beta} = 2\int_{\mathscr{S}}\mathrm{d}^3\zeta\,\rho \left\{  
\xi_{[\alpha} U_{\beta]} + \frac{1}{2} \xi_{[\alpha} U_{\beta]} \left( f_{\sigma\rho}\overset{*}{\xi}{}^\sigma \overset{*}{\xi}{}^\rho - R_{U\xi U\xi} \right) + \left( 1 + U\cdot \overset{*}{\xi} \right) \xi_{[\alpha} f_{\beta]\sigma} \overset{*}{\xi}{}^\sigma - \frac{2}{3} \xi_{[\alpha} R_{\beta]\xi U \xi} 
+ \mathcal{O}(\epsilon^4) \right\} \ .
\end{equation}
\end{widetext}
One can show by direct calculation that the spin $S_{\alpha\beta}$ and the momentum $P_\alpha$ from \eqref{Pfordust} satisfy the MPD equations \eqref{SteinhoffMPD2}. 

We now apply gauge conditions to analyze geodesic deviation. First, we use proper time to parametrize the fiducual worldline, so that $\sqrt{-\dot X^2} = 1$. With this choice we have $U^\alpha = \dot X^\alpha$ and $\overset{*}{\xi}{}^\alpha = \mathring{\xi}{}^\alpha$. Applying the proper time gauge \eqref{PT-gauge},  the momentum \eqref{pifordust} becomes
\begin{equation}\label{pideffordustgaugefixed}
\pi_\alpha  = \rho\left\{ \dot X_\alpha + \mathring{\xi}{}_\alpha 
- \frac{2}{3} R_{\dot X \xi \alpha \xi} + \mathcal{O}(\epsilon^3) \right\}
\ .
\end{equation}
The equation of motion  \eqref{pidoteqnfordust} reduces to 
\begin{eqnarray}\label{pidoteqnfordustgaugefixed}
\mathring{\pi}_\alpha & = & \rho \left\{ R_{\alpha \dot X\dot X\xi} + \frac{4}{3} R_{\alpha (\dot X\mathring{\xi})\xi} \right. \nonumber\\
& & \quad +\left.\frac{1}{3} R_{\alpha \dot X\dot X\xi;\xi} - \frac{1}{6}R_{\dot X\xi \dot X\xi;\alpha} +\mathcal{O}(\epsilon^3) \right\}  \ .\qquad
\end{eqnarray}
Finally, we can fix the fiducial worldline to be a geodesic; this implies $\mathring{U}{}^\alpha = \ringring{X}{}^\alpha = 0$. The 
equation of motion \eqref{pidoteqnfordustgaugefixed} together with the definition \eqref{pideffordustgaugefixed} yields 
\begin{align}
\begin{split}
\ringring{\xi}_\alpha &- \frac{2}{3} R_{\alpha\xi \dot X\xi;\dot X} + \frac{4}{3}R_{\alpha( \xi\mathring{\xi}) \dot X} \\
&= R_{\alpha \dot X\dot X\xi} + \frac{4}{3}R_{\alpha( \dot X\mathring{\xi})\xi} - \frac{1}{6}R_{ \dot X\xi \dot X\xi;\alpha}\\ 
&\quad + \frac{1}{3}R_{\alpha \dot X\dot X\xi;\xi} + \mathcal{O}(\epsilon^3) \ .
\end{split}
\end{align}
Applying the first and second Bianchi identities simplifies this to
\begin{equation}\label{second-order-geodesic}
\ringring{\xi}{}_\alpha = R_{\alpha \dot X\dot X\xi} + 2R_{\alpha\mathring{\xi} \dot X\xi} - \nabla_{(\xi}R_{\dot X)\alpha \dot X\xi}+ \mathcal{O}(\epsilon^3)
\end{equation}
which matches previous derivations of the geodesic deviation equations to second order \cite{Vines15,Aleksandrov}.


Geodesic deviation is often studied in terms of the $n$th order Jacobi fields $J^\alpha_n$, each of which is associated with a finite differential equation linear in $J^\alpha_n$ and only including lower-order Jacobi fields. We can generalize to our three-dimensional congruence of geodesics, representing Jacobi fields with spatial derivatives of the vector $\xi^\alpha$ evaluated at $\zeta_0^i$:
\begin{equation}
J^\alpha_{i_1\cdots i_n} \equiv \xi^\alpha_{,i_1\cdots i_n}(s,\zeta_0) \ .
\end{equation}
The equations of motion for the first- and second-order Jacobi fields {can be computed from} the first and second spatial derivatives of \eqref{second-order-geodesic}, respectively, {evaluated} at $\zeta_0^i$. This yields
\begin{subequations}
\begin{equation}
\ringring{J}{}^\alpha_i = R^{\alpha}{}_{ \dot{X}\dot{X}\beta}J^\beta_i \ ,
\end{equation}
\begin{align}
\begin{split}
\ringring{J}{}^\alpha_{ij} = &R^{\alpha}{}_{ \dot{X} \dot{X}\beta}J^\beta_{ij}+2\nabla_{(\beta}R^\alpha{}_{ \dot{X}) \dot{X}\gamma}J^\beta_{(i} J^\gamma_{j)} \\
&+ 4 R^\alpha{}_{\beta \dot{X}\gamma}\mathring{J}^\beta_{(i} J^\gamma_{j)}
\end{split}
\end{align}
\end{subequations}
which are the three-dimensional analogues {of} the traditional geodesic deviation equation and Bazanski's equation \cite{Bazanski}, respectively. These {results} are exact (requiring no further expansion in $\epsilon$) because all $\mathcal{O}(\epsilon^3)$ terms vanish when their first and second spatial derivatives are evaluated at $\zeta_0^i$.


Alternatively, instead of applying the proper time gauge and fixing the fiducial worldline to be a geodesic, we can choose the normal gauge $U_\alpha\xi^\alpha = 0$ and the center-of-mass condition $\Xi_\alpha = 0$.  For dust, the center of mass with density weight $w=\sqrt{h'}n^{\mu'}n^{\nu'}T_{\mu'\nu'}$ is 
\begin{equation}
\Xi_\alpha = \int_\mathscr{S}\mathrm{d}^3\zeta\,\gamma\rho\xi_\alpha
\end{equation}
where $\gamma = -n_{\mu'}U^{\mu'}$ is the relativistic gamma factor between the dust particles and the observers at rest in $\Sigma(s)$. The series expansion for $\gamma$ can be computed using Eqs.~\eqref{chainruledot},
\eqref{sigmaseries}, and the results for $H^{\mu'}{}_\alpha n_{\mu'}$ from 
\secref{conditions}. To second order, we find 
\begin{equation}
\gamma = 1+\frac{1}{2}f_{\alpha\beta}\overset{*}{\xi}{}^\alpha\overset{*}{\xi}{}^\beta+\mathcal{O}(\epsilon^3) \ .
\end{equation}
Under these gauge conditions, the spin becomes
\begin{widetext}
\begin{equation}
S_{\alpha\beta} = 2\int_\mathscr{S}\mathrm{d}^3\zeta\,\rho\left\{\left(1+U\cdot\overset{*}{\xi}\right)\xi_{[\alpha}f_{\beta]\gamma}\overset{*}{\xi}{}^\gamma - \frac{1}{2}R_{U\xi U\xi}\xi_{[\alpha}U_{\beta]}-\frac{2}{3}\xi_{[\alpha}R_{\beta]\xi U\xi}+\mathcal{O}(\epsilon^4)\right\} \ .
\end{equation}
\end{widetext}
To order $\epsilon^2$, the spin has the familiar form for the angular momentum of a system about its center of mass; roughly, $S_{\alpha\beta}$ is a sum over particles of the ``cross product'' between the position vector $\xi_\alpha$ and the relative momentum vector $\mathrm{d}^3\zeta\,\rho f_{\beta\gamma}\overset{*}{\xi}{}^\gamma$ (projected orthogonal to the four-velocity $U_\alpha$ of the fiducial worldline).

Taking the inner product of $S_{\alpha\beta}$ with $U^\beta$ yields
{ 
\begin{equation}
U^\beta S_{\beta\alpha} - \frac{1}{6}\int_\mathscr{S}\mathrm{d}^3\zeta\,\rho\,\xi_{\alpha}R_{U\xi U\xi}+\mathcal{O}(\epsilon^4) = 0  \ .
\end{equation}
This is precisely the center of mass condition $\Xi_\alpha =0$ with $\Xi_\alpha$ written in terms of spin as in Eq.~\eqref{SSC-correction}. It has the form of a spin supplementary condition with octupole correction.}

\section{Discussion}\label{discussion}
In \secref{mpd} we generalized the result of Steinhoff \cite{Steinhoff}, {to show} that ``effective particle'' actions of the form \eqref{EffectiveAction} always yield the MPD equations. These actions make no explicit reference to a body frame or angular velocity, and the spin is defined directly in terms of {the} internal structure of the effective particle. 

We define continuous bodies in terms of a congruence of world lines in \secref{continuous}, and show how  a continuous body can be expressed as an effective particle. The momentum and spin for this effective particle are exactly equivalent to Dixon's definitions for the momentum and spin of extended bodies.

{The MPD equations are often considered to be ``incomplete", requiring the addition of a spin supplementary condition. By starting with the action principle for a continuous body, written as an effective particle, we see clearly that the fiducial particle worldline is a gauge degree of freedom. The equations of motion, and the MPD equations in particular, are ``complete" in the sense of providing an unambiguous description of the evolution of the physical system. They are incomplete only in the sense that gauge conditions are needed to specify the evolution of the gauge variables. In \secref{conditions} we defined the center of mass of the continuous body and fixed the effective particle worldline to coincide with the center of mass. With the chosen center of mass density weight $w$, we showed that this gauge condition can be written as a spin--supplementary condition with higher--order corrections.}

{Note} that the action \eqref{Continuous-action} depends on the position $X^\mu$ through the metric tensor, which is treated as a fixed field. {Treating} the metric as a dynamical variable can provide a new direction for the study of gravitational self--force.

\section{Acknowledgments}\label{acknowledgements}
We would like to offer our gratitude to Justin Vines for sharing his notes and for helpful conversations. This research was supported by the Provost's Professional Experience Program at NCSU.
\begin{appendices}
\section{Covariant Variations and the Exponential Map}\label{tensor}
Consider a tensor $\psi_I$ that lives in the tangent space at a point 
$x$ in a manifold $\mathscr{M}$. The variation $\delta\psi_I = \psi'_I - \psi_I$ is the difference between $\psi_I$ and another tensor, $\psi'_I$, which is (in some 
sense) close to $\psi_I$. If $\psi'_I$ lives in a nearby tangent space at point $x'$, then $\delta\psi_I$ is not a tensor. We define the difference between $\psi'_I$ and $\psi_I$ as a tensor by parallel transporting $\psi'_I$ to the tangent space at $x$, then subtracting. This difference is 
$\Delta\psi_I = \delta\psi_I + \Gamma^\sigma_{\alpha\rho} \delta x^\alpha G^\rho{}_\sigma \psi_I$, 
where $\delta x = x' - x$. 
This leads us to define the covariant variation 
\begin{equation}\label{definedeltapsi}
	\Delta \equiv \delta + \Gamma^\sigma_{\alpha\rho} \delta x^\alpha G^\rho{}_\sigma
\end{equation}
acting on tensors. 
Here, $G^\rho{}_\sigma$ is an operator  acting on tensor indices. Specifically, $\mathrm{G}^\alpha{}_{\beta}$ are representations of the Lie algebra of the general linear group $\mathrm{GL}(4)$ acting on the tensor space \cite{DeWitt64}. In particular, for contravariant and covariant vectors $\psi^\mu$
and $\psi_\mu$, we have 
\begin{subequations}\label{Gactionexamples}
\begin{align}
	\mathrm{G}^\alpha{}_\beta \psi^\mu &= \delta_\beta^\mu \psi^\alpha \ ,\\
    \mathrm{G}^\alpha{}_\beta \psi_\mu &= -\delta^\alpha_\mu \psi_\beta \ .
\end{align}
\end{subequations}
The $G$'s satisfy the commutation relations
\begin{equation}\label{commutatorG}
[\mathrm{G}^{\alpha}{}_{\beta}, \mathrm{G}^{\mu}{}_{\nu}] = \delta^\mu_\beta \mathrm{G}^\alpha{}_{\nu} - \delta^\alpha_\nu \mathrm{G}^{\mu}{}_{\beta}
\end{equation}
for the Lie algebra of $\mathrm{GL}(4)$. 

For a given point $x\in\mathscr{M}$ and a vector $\xi\in T_{x}\mathscr{M}$, let $\gamma_{x,\xi}(u)$ be the affinely parameterized geodesic satisfying 
$\gamma_{x,\xi}(0)=x$ and $\dot{\gamma}_{x,\xi}(0)=\xi$. Define the exponential map as $\exp(x,\xi) = \gamma_{x,\xi}(1)$. In component form, we abbreviate $x^{\mu'} = \exp^{\mu'}(x,\xi)$, where $x^{\mu'}$ are the coordinates of the point $x'$. Primed indices are also used to denote tensors in the tangent space at $x'$. 

The variation of the exponential map is 
\begin{equation}
	\delta x^{\mu'} = \frac{\partial x^{\mu'}}{\partial x^\alpha}\delta x^\alpha + \frac{\partial x^{\mu'}}{\partial \xi^\alpha} \delta \xi^\alpha \ .
\end{equation}
As discussed above, $\delta\xi^\alpha$ is not a tensor whenever the variation of the base point $x$ is nonzero. Using the definition (\ref{definedeltapsi}), and Eq.~(\ref{Gactionexamples}), we have 
\begin{equation}
\delta x^{\mu'} = \left(\frac{\partial x^{\mu'}}{\partial x^\alpha} - \frac{\partial x^{\mu'}}{\partial \xi^\sigma} \Gamma_{\alpha\rho}^\sigma  \xi^\rho \right) \delta x^\alpha + \frac{\partial x^{\mu'}}{\partial \xi^\alpha} \Delta \xi^\alpha \ .
\end{equation}
The coefficients of $\delta x^\alpha$ and $\Delta\xi^\alpha$ are 
\begin{subequations}\label{HandK}
\begin{gather}\label{K}
K^{\mu'}{}_{\alpha}  \equiv \frac{\partial x^{\mu'}}{\partial x^\alpha} - \Gamma^\gamma_{\alpha\beta}\xi^\beta\frac{\partial x^{\mu'}}{\partial \xi^\gamma} \ ,\\\label{H}
H^{\mu'}{}_{\alpha} \equiv \frac{\partial x^{\mu'}}{\partial \xi^\alpha} \ .
\end{gather}
\end{subequations}
These are bitensors, meaning that they depend on two points $x$ and $x'$, and they have indices that exist in the tangent spaces of both points. 

Using this notation, the variation of the exponential map becomes
\begin{equation}\label{deltaxprime}
	\delta x^{\mu'} = K^{\mu'}{}_\alpha \delta x^\alpha + H^{\mu'}{}_\alpha  \Delta \xi^\alpha \ .
\end{equation}
We use this result in the main text to compute results (\ref{chainruledot}) 
and (\ref{chainrulei}). 

The definitions of $H^{\mu'}{}_{\,\alpha}$ and $K^{\mu'}{}_{\, \alpha}$ as the vertical and horizontal derivatives of the exponential function were, to our knowledge, first given in \cite{Vines15}, though Dixon employs similar definitions in \cite{Dixon15}. $H$ and $K$ are often called the {\em Jacobi propagators} and were first discovered in the context of Synge's world function $\sigma$ \cite{Dixon70-1}.

Next, consider a tensor function $T(x,\psi_I)$ on the tensor bundle. That is, $T(x,\psi_I)$ depends on the point $x\in\mathscr{M}$ and tensor arguments $\psi_I$. Both $T$ and $\psi_I$ exist in the tangent space of $x$. The variation of $T$ is
\begin{equation}\label{deltaT}
\delta T = \frac{\partial T}{\partial x^\alpha}\delta x^\alpha +
\frac{\partial T}{\partial \psi_I}\delta\psi_I \ .
\end{equation}
When $\delta x^\alpha$ is not zero, neither $\delta T$ nor $\delta \psi_I$ are tensors.  Using the covariant variation (\ref{definedeltapsi}), we have 
\begin{align}\label{coT}
\begin{split}
\Delta T = &\left(\frac{\partial T}{\partial x^\alpha}+\Gamma^\sigma_{\alpha\rho}\mathrm{G}^{\rho}{}_{\sigma}T - \frac{\partial T}{\partial\psi_I} \Gamma^\sigma_{\alpha\rho} \mathrm{G}^\rho{}_\sigma \psi_I \right)\delta x^\alpha \\
&+ \frac{\partial T}{\partial \psi_I}\Delta\psi_I \ .
\end{split}
\end{align}
Each of the factors on the right--hand side of this equation is covariant. Note that the 
first two terms in parenthesis give the traditional definition of the covariant derivative,
\begin{equation}\label{coderiv}
\nabla_\alpha T \equiv \frac{\partial T}{\partial x^\alpha}+\Gamma^\sigma_{\alpha\rho}\mathrm{G}^{\rho}{}_{\sigma}T \ .
\end{equation}
When $T$ depends on $\psi_I$, $\nabla_\alpha T$ is not covariant. Covariance requires the third term in parenthesis above, which arises from the parallel transport of $\psi$ as the base point $x$ is varied.

It would be a misnomer to call $\nabla_\alpha$ the ``covariant derivative": we shall opt here to call it by its other traditional name, the Levi-Civita derivative. The covariant derivative is defined by 
\begin{equation}
\frac{\mathcal{D}T}{\mathcal{D} x^\alpha} \equiv \nabla_\alpha T - \frac{\partial T}{\partial \psi_I} \Gamma^\sigma_{\alpha\rho} \mathrm{G}^\rho{}_\sigma \psi_I \ .
\end{equation}
With this notation, the covariant variation of $T$ becomes
\begin{equation}\label{covariantvariation}
\Delta T = \frac{\mathcal{D}T}{\mathcal{D} x^\alpha} \delta x^\alpha + \frac{\partial T}{\partial \psi_I} 
\Delta\psi_I \ .
\end{equation}
We also use the shorthand notation  $\mathcal{D}_\alpha T = \mathcal{D}T/\mathcal{D}x^\alpha$. It is worth noting that when $\psi_I$ is the contravariant vector $\xi^\alpha$, the covariant derivative $\mathcal{D}_\alpha$ corresponds to the vertical derivative $\nabla_{\ast\alpha}$ defined by Dixon and $\partial/\partial\psi_I=\partial/\partial\xi^\alpha$ corresponds to the horizontal  derivative $\nabla_{\alpha\ast}$ \cite{Dixon74,Dixon79}. The notation  $\mathcal{D}T/\mathcal{D}x^\alpha$ for the   covariant partial derivative is adopted from Vines \cite{Vines16}. 

For the case where $T$ has no explicit dependence on $x$ (that is, $T=T(\psi_I)$), Eq.~\eqref{deltaT} becomes 
$\delta T = (\partial T/\partial\psi_I)\delta\psi_I$. 
We can derive a ``chain rule" for the generators $\mathrm{G}^\rho{}_{\sigma}$ by considering the passive coordinate transformations $\tilde{x}^\alpha = x^\alpha+\delta x^\alpha$. Note that, in this context, $\delta x^\alpha$ denotes the difference between coordinate values, using two different coordinate systems, at a single point in the manifold $\mathcal{M}$.  Elsewhere in this appendix, $\delta x^\alpha$ denotes the difference between coordinates values, using a single coordinate system, for two distinct points of $\mathcal{M}$. 

Under the passive coordinate transformation, $T$ and $\psi_I$ transform as
\begin{subequations}
\begin{gather}
\delta T = -\partial_\alpha \delta x^\beta \mathrm{G}^\alpha{}_\beta T \ ,\\
\delta \psi_I = -\partial_\alpha \delta x^\beta \mathrm{G}^\alpha{}_\beta \psi_I \ .
\end{gather}
\end{subequations}
Using these results in the relation $\delta T = (\partial T/\partial\psi_I)\delta\psi_I$ and noting that $\partial_\alpha \delta x^\beta$ is arbitrary, we have
\begin{equation}\label{specialchainrule}
\mathrm{G}^\alpha{}_\beta T = \frac{\partial T}{\partial \psi_I} 
\mathrm{G}^\alpha{}_\beta\psi_I 
\end{equation}
In this paper we often consider the case $T=T(\phi_A,\psi_I)$, where $\phi_A=\phi_A(x)$ is a field defined on the manifold, and $\psi_I(s)$ is a tensor function defined solely along a world line {$x^\alpha = X^\alpha(s)$.} In this case the results \eqref{coT} and \eqref{specialchainrule} become
\begin{subequations}\label{covvarall}
\begin{eqnarray}\label{covvarPhi}
\Delta T & = & \frac{\partial T}{\partial \phi_A} \Delta\phi_A +
\frac{\partial T}{\partial \psi_I}\Delta\psi_I  \ ,\\
\label{ChainG}
\mathrm{G}^\alpha{}_\beta T & = & \frac{\partial T}{\partial \phi_A}\mathrm{G}^\alpha{}_\beta \phi_A +
\frac{\partial T}{\partial \psi_I}\mathrm{G}^\alpha{}_\beta\psi_I \ ,
\end{eqnarray}
\end{subequations}
respectively.

The results (\ref{covvarall}) are applied in the main text to the effective particle Lagrangian, which is a spacetime scalar, by evaluating these relations 
at the spacetime point $x = X(s)$ on the particle worldline. In particular,
the variation of the action Eq.~\eqref{EffectiveVar1} and the property 
\eqref{chainruleapplied} of the Lagrangian are obtained from Eqs.~(\ref{covvarall}). 
The results (\ref{covvarall}) are also applied to the Lagrangian density for the continuous 
body by evaluating them at the spacetime point $x = X(s,\zeta)$. See for example the variation of the action in Eq.~(\ref{deltaSforcontinuum}) and the calculation leading to the stress--energy tensor \eqref{ContinuousStressEnergy}.

Now suppose that $T_1(x,\psi_I)$ is a tensor function on the tensor {bundle} defined by $T_1(x,\psi_I) \equiv T_2(\phi_A(x,\psi_I),\psi_I)$, where 
$\phi_A(x,\psi)$ are also functions on the tensor bundle. 
By applying Eq.~\eqref{covvarPhi} to $T_2$ and Eq.~\eqref{covariantvariation} to $T_1$ and $\phi_A$, we find 
the chain rules
\begin{subequations}\label{bundle-chain}
\begin{align}
\frac{\mathcal{D}T_1}{\mathcal{D}x^\alpha} &= \frac{\partial T_2}{\partial \phi_A}\frac{\mathcal{D}\phi_A}{\mathcal{D}x^\alpha}\\
\frac{\partial T_1}{\partial \psi_I} &= \frac{\partial T_2}{\partial \phi_A}\frac{\partial \phi_A}{\partial \psi_I} + \frac{\partial T_2}{\partial \psi_I} \ .
\end{align}
\end{subequations}
These results are used in Appendix \ref{bitensor}.

Finally, suppose we have two independent variations $\Delta_1$ and $\Delta_2$ acting on a general tensor $T(x,\psi_I)$. The covariant variation $\Delta_1$ corresponds to a change in $x^\alpha$ by $\delta_1 x^\alpha$ and $\Delta_2$ corresponds to the change $\delta_2 x^\alpha$. Explicitly, we have  $\Delta_1 = \delta_1 + \Gamma^\sigma_{\alpha\rho} \delta_1 x^\alpha$ and $\Delta_2 = \delta_2 + \Gamma^\sigma_{\alpha\rho} \delta_2 x^\alpha$.  In general these variations will not commute. Rather,
\begin{align}
\begin{split}
[\Delta_1,\Delta_2] &= [\delta_1 + \Gamma^\sigma_{\alpha\rho}\delta_1 x^\alpha \mathrm{G}^{\rho}{}_{\sigma}, \Delta_2 ]\\
&= \delta_1\Delta_2 + \Gamma^\sigma_{\alpha\rho} \delta_1 x^\alpha \mathrm{G}^\rho{}_\sigma \Delta_2 - (1\leftrightarrow 2)
\end{split}
\end{align}
where $(1\leftrightarrow 2)$ denotes the previous terms with indices 
$1$ and $2$ exchanged. Note that $\Delta_2$ is a scalar operator---it carries no tensor indices---so it commutes with the generator $\mathrm{G}^\rho{}_\sigma$. Thus, we have
\begin{align}
\begin{split}
\mathrm{G}^\rho{}_\sigma \Delta_2 T &= \Delta_2 ( \mathrm{G}^\rho{}_\sigma T) \\
&= \delta_2 (\mathrm{G}^\rho{}_\sigma T) + \Gamma_{\beta\mu}^\nu
\delta_2 x^\beta \mathrm{G}^\mu{}_\nu \mathrm{G}^\rho{}_\sigma T \ . \label{GandDeltaeqnOLD}
\end{split}
\end{align}
With this result it is straightforward to obtain
\begin{align}
\begin{split}
[\Delta_1,\Delta_2] &= \delta_1\delta_2 + \delta_1(\Gamma_{\beta\mu}^\nu 
\delta_2 x^\beta \mathrm{G}^\mu{}_\nu)\\ 
&+ \Gamma_{\alpha\rho}^\sigma \Gamma_{\beta\mu}^\nu \delta_1 x^\alpha \delta_2 x^\beta \mathrm{G}^\mu{}_\nu \mathrm{G}^\rho_{\sigma} - (1\leftrightarrow 2)  \ .
\end{split}
\end{align}
(The variation $\delta_1$ in the second term on the right acts only on the factor 
in parenthesis.) Terms such as $\delta_1\Gamma^\nu_{\beta\mu}$ yield derivatives of 
Christoffel symbols. The terms  $ \mathrm{G}^\mu{}_\nu \mathrm{G}^\rho_{\sigma}$ are simplified using 
the commutation relations (\ref{commutatorG}). The final result is 
\begin{align}
\begin{split}\label{DeltaDeltaCommutator}
[\Delta_1,\Delta_2] = [\delta_1,\delta_2] &+ \Gamma_{\alpha\mu}^\nu  ([\delta_1,\delta_2]x^\alpha)\mathrm{G}^\mu{}_\nu\\ 
&+ R^\sigma{}_{\rho\mu\nu} \delta_1x^\mu \delta_2x^\nu \mathrm{G}^\rho{}_\sigma \ ,
\end{split}
\end{align}
where $R^\sigma{}_{\rho\mu\nu}$ is the Riemann tensor. Note that the first two terms on the 
right--hand side are the covariant extension of the operator $[\delta_1,\delta_2]$. 

Now suppose $x$ is a point on a worldline, $x = X(s)$, so the tensors $T$, $\phi_A$ and $\psi_I$ depend on the parameter $s$. Let $\delta_2 = \delta s(\mathrm{d}/\mathrm{d} s)$, where $\delta s$ is infinitesimal. Then $\delta_2 x^\alpha = \dot X^\alpha\delta s$ and $\Delta_2 = \delta s (\mathrm{D}/\mathrm{D}s)$, 
where $\mathrm{D}/\mathrm{D}s$ is the covariant derivative along the worldline. We can also drop the subscripts from $\delta_1$ and $\Delta_1$. The result (\ref{DeltaDeltaCommutator}) now gives the
commutator of a general covariant variation $\Delta$ with the covariant 
derivative $\mathrm{D}/\mathrm{D}s$. Since $\delta$ 
and $\mathrm{d}/\mathrm{d}s$ commute, we have
\begin{equation}\label{commutatorDeltaD}
\left[\Delta, \frac{\mathrm{D}}{\mathrm{D}s}\right] = R^\sigma{}_{\rho\alpha\beta}\delta X^\alpha \dot{X}^\beta \mathrm{G}^\rho{}_{\sigma} \ .
\end{equation}
This result is used in the variation of the Lagrangian in \secref{mpd}.

\section{Bitensors and Synge's World Function}\label{bitensor}
Synge's world function  is closely related to the exponential function. Given two points $x$ and $x'$, take the geodesic $\gamma_{x,x'}(u)$ affinely parametrized so that $\gamma_{x,x'}(0)=x$ and $\gamma_{x,x'}(1)=x'$.
Then Synge's world function is defined as \cite{Synge}
\begin{equation}\label{synge}
\sigma(x,x') \equiv \frac{1}{2}\int_{0}^{1}\mathrm{d}u\ g_{\alpha\beta}\left(\gamma_{x,x'}(u)\right)\frac{\mathrm{d}\gamma_{x,x'}^\alpha}{\mathrm{d}u}\frac{\mathrm{d}\gamma_{x,x'}^\beta}{\mathrm{d}u} \ ,
\end{equation}
which is one-half the squared distance of the geodesic connecting $x$ and $x'$. Obviously, this function is well-defined  only when $x$ and $x'$ are close enough that they are  connected by one geodesic.
In this paper we consider the approximation of small bodies and assume that this is the case for any spacelike--separated $x$ and $x'$ in the world tube $\mathscr{W}$ of our body.

When taking derivatives of $\sigma$, we omit semicolons following the standard notation, and as before use primed indices for derivatives with respect to $x'$
and unprimed indices for $x$. It can be seen that Synge's world function satisfies the formulae \cite{Poisson}
\begin{subequations}
\begin{gather}\label{syngetangent1}
\sigma^\alpha(x,x') = -\dot{\gamma}_{x,x'}^\alpha(0) \ ,\\ \label{syngetangent2}
\sigma^{\mu'}(x,x') = \dot{\gamma}_{x,x'}^{\mu'}(1) \ ,\\\label{syngenorm}
\sigma^{\mu'}\sigma_{\mu'} = \sigma^\alpha\sigma_\alpha = 2\sigma \ ,\\\label{syngecontraction1}
\sigma^{\mu'}_\alpha \sigma_{\mu'} = \sigma^\beta_\alpha\sigma_\beta = \sigma_\alpha \ ,\\\label{syngecontraction2}
\sigma^{\nu'}_{\mu'}\sigma_{\nu'} = \sigma^{\alpha}_{\mu'}\sigma_\alpha = \sigma_{\mu'} \ .
\end{gather}
\end{subequations}

Based on \eqref{syngetangent1} and the definition of the exponential map, we see that $x'(x,\xi)$ and $\sigma(x,x')$ are related by
\begin{equation}\label{syngetoexp}
\xi^\alpha = -\sigma^\alpha\left(x,x'\right) \ .
\end{equation}
We can treat each side of this equation as a tensor function of $x^\alpha$ and $\xi^\alpha$ by writing $\xi^\alpha = -\sigma^\alpha(x,x'(x,\xi))$. This allows us to apply the covariant variation (\ref{covariantvariation}) and get
\begin{equation}\label{syngevariation}
\Delta \xi^\alpha = -\left(\sigma^\alpha{}_{\beta} + \sigma^\alpha{}_{\mu'}K^{\mu'}{}_{\beta}\right)\delta x^\beta -\sigma^\alpha{}_{\mu'} H^{\mu'}{}_{\beta}\Delta \xi^\beta
\end{equation}
where we have used the chain rule \eqref{bundle-chain}. Matching coefficients gives $-\sigma^\alpha_{\mu'} H^{\mu'}{}_{\beta}=\delta^\alpha_\beta$ and $\sigma^\alpha{}_{\beta} + \sigma^\alpha{}_{\mu'}K^{\mu'}{}_{\beta}=0$, which imply
\begin{subequations}\label{DixonHandK}
\begin{align}\label{DixonK}
K^{\mu'}{}_{\alpha} &= -\overset{-1}{\sigma}{}^{\mu'}{}_\beta\sigma^{\beta}{}_{\alpha} \ ,\\\label{DixonH}
H^{\mu'}{}_{\alpha}&= -\overset{-1}{\sigma}{}^{\mu'}{}_{\alpha} \ .
\end{align}
\end{subequations}
{The} symbol $-1$ over the matrices indicates matrix inversion. Equations \eqref{DixonH} and \eqref{DixonK} are Dixon's original definitions for $H$ and $K$ \cite{Dixon70-1}.

We can use the Jacobi propagator $H$ and its inverse to transfer indices on bitensors into a single tangent space, so that we can sensibly define series expansions for bitensors. In Riemann normal coordinates, the components of $\xi^\alpha(x,x')$ are (by definition) equal to the coordinates of the point $x'$. Thus, in Riemann normal coordinates, we have $H^{\mu'}{}_{\alpha} = \delta^{\mu'}_{\alpha}$. (This equality is by components, and is obviously not a tensor equation.) As such, the coefficients of a series expansion mediated by $H$ and its inverse can be found from the Taylor expansion of the same tensor field in Riemann normal coordinates. For example, in this paper we use the following expansion for the metric tensor and its inverse:
\begin{subequations}\label{g-exp}
\begin{align}
\begin{split}
H^{\mu'}{}_{\alpha}H^{\nu'}{}_\beta g_{\mu'\nu'}(x') =& g_{\alpha\beta}(x) - \frac{1}{3}R_{\alpha\xi\beta\xi}(x)\\ 
&- \frac{1}{6} R_{\alpha\xi\beta\xi;\xi}(x) + \mathcal{O}(\epsilon^4) \ ,\label{g-expHH}
\end{split}\\
\begin{split}
\overset{-1}{H}{}^{\alpha}{}_{\mu'}\overset{-1}{H}{}^{\beta}{}_{\nu'} g^{\mu'\nu'}(x') =& g^{\alpha\beta}(x) + \frac{1}{3}R^\alpha{}_{\xi}{}^{\beta}{}_{\xi}(x)\\ 
&+ \frac{1}{6}R^\alpha{}_{\xi}{}^{\beta}{}_{\xi;\xi}(x) + \mathcal{O}(\epsilon^4) \ .\label{g-expHinvHinv}
\end{split}
\end{align}
\end{subequations}
These follow from the familiar expansions for the metric and its inverse in Riemann normal coordinates \cite{Brewin09}. Here, $\xi$ appears as an index in the Riemann curvature tensor, indicating contraction of that index with $\xi^\alpha$. The coefficients in the $H$-series expansion of a tensor field are the {\em Veblen extensions} of that tensor field \cite{Veblen}.

Another important bitensor is the parallel propagator $g^{\mu'}{}_\alpha(x,x')$. This bitensor parallel propagates contravariant indices from $T_x\mathscr{M}$ to $T_{x'}\mathscr{M}$ along the geodesic $\gamma_{x,x'}$. Because the metric is parallel propagated, we have $g^{\alpha\beta}(x)g^{\mu'}{}_\alpha g^{\nu'}{}_{\beta} = g^{\mu'\nu'}(x')$. It follows that $g^\alpha{}_{\mu'}$ parallel propagates covariant indices from $T_x\mathscr{M}$ to $T_{x'}\mathscr{M}$. We also have the relations $g^\alpha{}_{\mu'}g^{\mu'}{}_{\beta}=\delta^\alpha_\beta$ and $g^{\mu'}{}_{\alpha}g^{\alpha}{}_{\nu'}=\delta^{\mu'}_{\nu'}$, so that $g^{\alpha}{}_{\mu'}$ and $g^{\mu'}{}_{\alpha}$ can inversely propagate contravariant and covariant indices from $T_{x'}\mathscr{M}$ to $T_x\mathscr{M}$, respectively.

We can also use the parallel propagator to define series expansions for bitensors. In this paper we use the expansions \cite{Vines16}
\begin{subequations}\label{HK-exp}
\begin{gather}
g^\alpha{}_{\mu'}H^{\mu'}{}_{\beta} = \delta^\alpha_\beta - \frac{1}{6}R^\alpha{}_{\xi\beta\xi} - \frac{1}{12} R^\alpha{}_{\xi\beta\xi;\xi}+\mathcal{O}(\epsilon^4),\\
g^\alpha{}_{\mu'}K^{\mu'}{}_{\beta} = \delta^\alpha_\beta - \frac{1}{2}R^\alpha{}_{\xi\beta\xi} - \frac{1}{6} R^\alpha{}_{\xi\beta\xi;\xi}+\mathcal{O}(\epsilon^4) 
\end{gather}
\end{subequations}
for the Jacobi propagators. 


\end{appendices}
\bibliography{references}
\bibliographystyle{unsrt}
\end{document}